\documentclass[12pt, journal, draftclsnofoot, onecolumn]{IEEEtran}
\usepackage{setspace}
\usepackage{lineno}
\usepackage{amsmath,amssymb,amsfonts}
\usepackage{algorithmic}
\usepackage{algorithm}
\usepackage{array}
\usepackage[caption=false,font=normalsize,labelfont=sf,textfont=sf]{subfig}
\usepackage{textcomp}
\usepackage{stfloats}
\usepackage[english]{babel}
\usepackage{url}
\usepackage{verbatim}
\usepackage{physics} 
\usepackage{mathtools}
\usepackage{graphicx}
\usepackage{xcolor}
\usepackage{cite}
\usepackage{booktabs} 
\def\BibTeX{{\rm B\kern-.05em{\sc i\kern-.025em b}\kern-.08em
    T\kern-.1667em\lower.7ex\hbox{E}\kern-.125emX}}
\usepackage{balance}
\usepackage{hyperref}
\usepackage{orcidlink}

\hypersetup{
    colorlinks=true,      
    linkcolor=blue,       
    citecolor=blue,       
    urlcolor=blue         
}
\usepackage{ragged2e}

\begin{document}
\title{Performance Analysis of Underwater Quantum Key Distribution Protocols: BB84, SARG04, and BBM92}

\author{
\IEEEauthorblockN{
Nour Rizk \textsuperscript{\Large\orcidlink{0009-0009-0171-2388}},
Angélique Drémeau \textsuperscript{\Large\orcidlink{0000-0001-6325-3695}}, and
Arnaud Coatanhay \textsuperscript{\Large\orcidlink{0000-0003-0101-3105}}~\IEEEmembership{}
}
\thanks{N. Rizk, A. Drémeau and A. Coatanhay are with Lab-STICC, UMR CNRS 6285, ENSTA, IP Paris, 29806 Brest, France.}
}



\maketitle

\begin{abstract}
This study compares the quantum bit error rate (QBER) performance of
BB84, SARG04, and the entanglement-based BBM92 protocol in non-turbulent
underwater optical channels. Clear, coastal, and turbid water types are
considered under different background-illumination and receiver
configurations. For BB84 and SARG04, channel attenuation and
background-induced detections determine the quantum gain and QBER. For
BBM92, photon loss is included through the two arm detection efficiencies,
whereas the polarization state conditioned on two-photon detection is
described by local depolarizing Kraus channels. The BBM92 QBER is obtained
from the correlations in both measurement bases and from an exact partition
of recorded events into true and false coincidences. Stochastic numerical
estimates provide consistency checks of the analytical implementation.
The results show that water turbidity and optical background reduce the
distance below the adopted QBER threshold. They also show that a low
conditional BBM92 QBER may coexist with a very small coincidence
probability; therefore, the QBER-threshold distance alone does not represent
a complete operational or secret-key-rate bound.
\end{abstract}


\section{Introduction}
\IEEEPARstart{U}{nderwater} communication links between submarines,
autonomous vehicles, and sensor networks may carry sensitive information.
Underwater optical wireless communication (UOWC) offers high bandwidth and
low latency compared with acoustic communication
\cite{theocharidis2025underwater,kaushal2016underwater,
lanzagorta2012underwater}. Its useful range is nevertheless strongly
limited by absorption and scattering in seawater \cite{baykal2022underwater}.

{
Under the assumptions of a specified protocol, authenticated classical
communication, and sufficiently characterized devices, QKD can provide
information-theoretic key security grounded in quantum mechanics
\cite{busch2007heisenberg,wootters1982single}. Eavesdropping is inferred
statistically from disturbances observed during parameter estimation, rather than
detected deterministically on every signal. Since its first 32-cm
demonstration \cite{bennett1989experimental}, QKD has progressed across fiber,
free-space, and satellite links
\cite{takesue200610,takesue2007quantum,wang20122,buttler1998practical,
sharma2015controlled,sidhu2021advances,ecker2021strategies,
acosta2024analysis}. Applying QKD to underwater environments, however, requires
overcoming severe optical constraints, including absorption, scattering, ambient
background illumination, and polarization degradation
\cite{ji2017towards,zhao2019performance}.
}

The performance of QKD protocols is commonly characterized by the quantum
bit error rate (QBER), defined as the fraction of erroneous outcomes among
the retained detection events. It provides an indicator of channel
degradation and may also reveal disturbances compatible with an
eavesdropping attempt.
The first QKD protocol, BB84, was introduced by Bennett and Brassard \mbox{in 1984 \cite{brassard1984quantum}.} It is based on single-photon emission in four polarization states belonging to two orthonormal bases. It offers a low QBER and long transmission distances for an error threshold of $11\%$ \cite{shor2000simple}, but these results remain \mbox{theoretical \cite{muskan2023analysing}.} The first experimental demonstration of free-space underwater QKD was performed in 2017 \mbox{over 3.3 m \cite{ji2017towards},} followed by a successful implementation of the BB84 protocol in 2019 \mbox{over 2.37 m \cite{zhao2019performance}.} Recently, research has explored BB84 under the effects of turbulence, temperature, and salinity, as well as in different water \mbox{conditions \cite{ata2025impact, paglierani2023primer}.}

{
Although BB84 is widely implemented, practical source and detector
imperfections motivate the investigation of alternative protocols. SARG04 was
introduced with a modified state-announcement procedure to improve resilience
against multiphoton attacks \cite{scarani2004quantum}. In this work, we evaluate
and compare BB84 and SARG04 under a comprehensive underwater link model that
explicitly integrates realistic Poisson photon-number statistics, channel
transmittance, and ambient-background detections to establish benchmark
QBER-threshold distances across clear, coastal, and turbid waters
\cite{rizkperformances}.

Beyond prepare-and-measure protocols, entanglement-based schemes such as BBM92
\cite{bennett1992quantum} offer distinct operational and physical dynamics. In a
two-arm BBM92 configuration, a central source distributes entangled photon pairs
to Alice and Bob. While existing underwater QKD studies focus primarily on
prepare-and-measure architectures, modeling entanglement distribution in aquatic
media requires an analytical framework capable of capturing both optical
attenuation and quantum state decoherence.

To address this, we develop a modeling framework for underwater BBM92 that
explicitly decouples optical loss from conditional polarization degradation. By
combining local depolarizing Kraus channels with an exact partition of recorded
events into true and false coincidences, our model reveals the interplay
between surviving two-photon polarization correlations and background-induced
accidental coincidences. Furthermore, we systematically investigate the impact of
the entangled-source position, demonstrating how spatial asymmetry influences the
achievable QBER performance in various underwater noise regimes.
}

{
Specifically, the main contributions of this article are threefold:
first, we provide a unified comparative analysis of BB84, SARG04, and BBM92 QBER
performance in non-turbulent underwater optical channels across diverse water
turbidity and ambient-illumination scenarios.
Second, we develop an exact coincidence-based QBER framework for BBM92 that
isolates the reduction in detected-pair rate from the Kraus-modeled depolarization
of conditional two-photon states, avoiding any double-counting of optical
attenuation.
Third, we analyze the role of the source position in two-arm BBM92 links and
determine the conditions under which symmetric placement optimizes the
QBER-threshold distance.
Finally, event-based stochastic Monte Carlo simulations are implemented to
verify the mathematical and numerical consistency of the proposed analytical models.
}

The remainder of this article is organized as follows. 
Section~\ref{System model} describes the system model and provides a detailed description of the protocols. 
The impact of the underwater channel on quantum transmission is analyzed in Section~\ref{Impact of the underwater channel on quantum transmission}. 
Section~\ref{Performance metrix for UQKD Communication system} presents the analysis of quantum gain and QBER. 
The evolution of entangled states using Kraus operators in an underwater quantum channel is detailed in Section~\ref{Modeling Entangled States with Kraus Operators in an Underwater Quantum Channel}. 
Section~\ref{Results} discusses the obtained results. 
Finally, the conclusions are presented in Section~\ref{Conclusion}.

\section{System Model}
\label{System model}
\begin{figure}[!t]
\centering
\includegraphics[width=1.0\columnwidth]{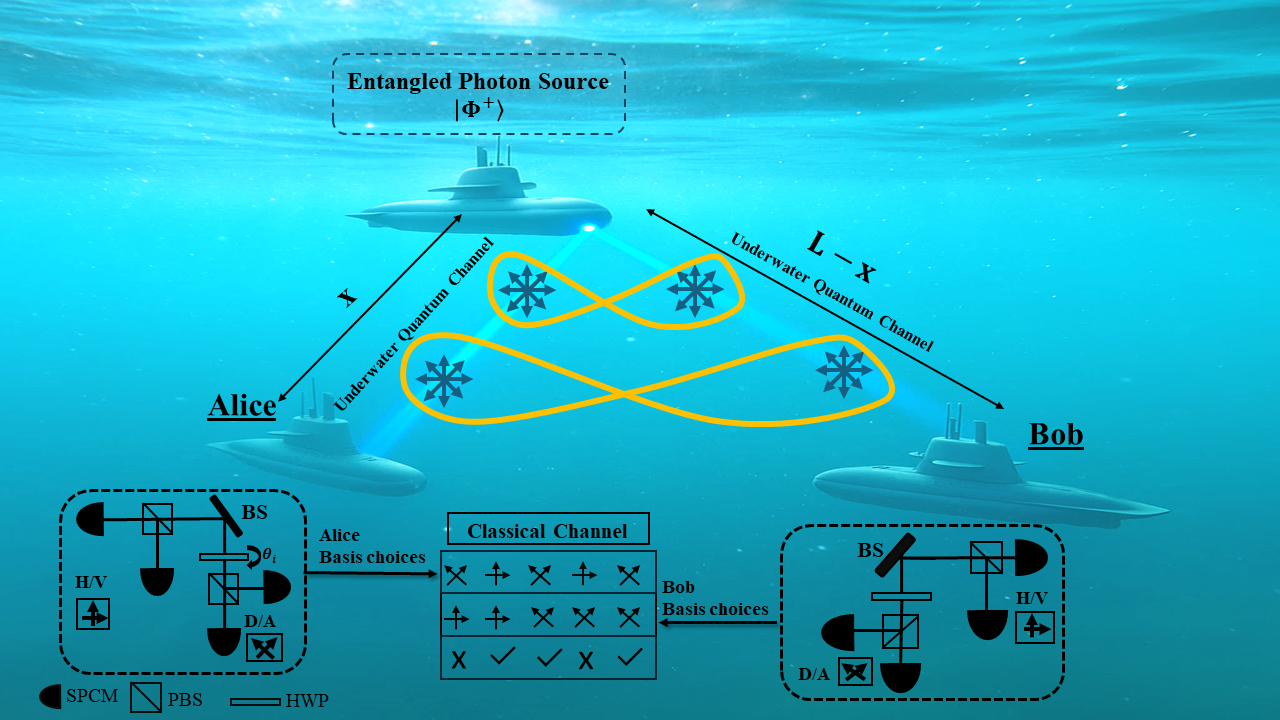}
\caption{{Schematic diagram of the entanglement-based BBM92 QKD configuration considered in this work. A central entangled-photon source distributes one photon to Alice over the optical path \(x\) and the other photon to Bob over the optical path \(L-x\), where \(L\) denotes the total two-arm propagation length. The half-wave plates (HWP) set the measurement-basis orientations \(\theta_i\); BS, PBS, and SPCM denote beam splitter, polarizing beam splitter, and single-photon counting module, respectively. The prepare-and-measure BB84 and SARG04 protocols are discussed separately as benchmark configurations.}}
\label{figure system model}
\end{figure}

{
In the BBM92 configuration considered here, a source generates
polarization-entangled photon pairs and distributes one photon to Alice over
a path of length \(x\) and the other to Bob over a path of length \(L-x\)
through a non-turbulent underwater optical channel, as illustrated in
Fig.~\ref{figure system model}.}

{
When the entangled-photon source is located at Alice (\(x=0\)), the BBM92
link becomes strongly asymmetric: Alice detects one photon locally, while
the second photon propagates over the full distance \(L\) to Bob. This
configuration remains entanglement based and is therefore not equivalent
to the prepare-and-measure configurations used by BB84 and SARG04.
}
{
For each detection window, Alice and Bob independently choose the
\(H/V\) or \(D/A\) basis. Attenuation reduces the number of recorded pairs,
whereas polarization degradation may alter their conditional correlations.
They retain same-basis outcomes as sifted data after announcing their basis
choices over an authenticated classical channel. Secret-key extraction
would additionally require parameter estimation, error correction, and
privacy amplification, which are not modeled here.
}
{
Here, \(\ket{H}\) and \(\ket{V}\) denote horizontal and vertical linear
polarizations, while the diagonal \(\ket{D}\) and antidiagonal \(\ket{A}\) states are
\[
\ket{D}
=
\frac{\ket{H}+\ket{V}}{\sqrt{2}},
\qquad
\ket{A}
=
\frac{\ket{H}-\ket{V}}{\sqrt{2}}.
\]
In the SARG04 notation used below, we set $\ket{+}\equiv\ket{D}$ and $\ket{-}\equiv\ket{A}$ to avoid confusion with the identifier $A$ used for Alice.
}

{
Changing the BBM92 source position modifies the two propagation lengths,
the arm detection efficiencies, and the conditional polarization
correlations. The BBM92 measurement and classical post-processing rules
remain unchanged. BB84 and SARG04 are treated separately as
prepare-and-measure benchmark configurations.
}

\subsection{BB84 protocol}
{
The BB84 protocol, introduced by Bennett and Brassard in 1984
\cite{brassard1984quantum}, is a prepare-and-measure protocol in which the
source is located at Alice. The source-position variable \(x\), introduced
for the two-arm BBM92 geometry, is therefore not used for BB84. Alice
prepares polarization states in the \(H/V\) or \(D/A\) basis and sends them
to Bob. After transmission and measurement, Alice and Bob announce their
basis choices over an authenticated classical channel and retain only
same-basis outcomes. A random subset is disclosed for QBER estimation,
defined as
}
\begin{equation}
\label{QBER_BB84_simulé}
\text{QBER}_\text{BB84} = \frac{\text{Number of erroneous sifted bits}}{\text{Total number of sifted bits}}.
\end{equation}
{
The value \(11\%\) is used as an asymptotic QBER reference from
\cite{shor2000simple}. It defines a performance threshold under the cited
model, not a complete implementation-security statement; protocols should
not be ranked by this threshold alone \cite{fung2005performance}.
}

\subsection{SARG04 protocol}
SARG04 protocol, proposed by Scarani et al. in 2004 \cite{scarani2004quantum}, was designed to enhance the security of QKD against PNS attacks, which exploit multiphoton emissions from weak coherent sources. {As in BB84, the source is located at Alice; the BBM92 source-position variable \(x\) is not used in this prepare-and-measure configuration.} The key difference lies in the sifting process: instead of directly revealing the measurement basis, Alice publicly announces a set of two non-orthogonal states chosen from $\{ \ket{H},\ket{+}\}\,, \, \{ \ket{H},\ket{-}\}\, ,\, \{ \ket{V},\ket{+}\}\, ,\, \{ \ket{V},\ket{-}\}$. Bob then compares his measurement outcome with the announced set. If his result is orthogonal to one of the two states, he can conclude that the photon corresponds to the other state, the bit is therefore conclusive. Otherwise, if his result is compatible with both states, it is inconclusive and discarded. The QBER is defined over the conclusive bits as
\begin{equation}
\label{QBER_SARG04_simulé}
\text{QBER}_\text{SARG04} = \frac{\text{Number of erroneous conclusive bits}}{\text{Total number of conclusive bits}}.
\end{equation}
{
The value \(14.9\%\) is used as the asymptotic SARG04 reference threshold
from \cite{branciard2005security} and only defines a QBER-threshold
distance here. SARG04 can improve resistance to some photon-number-splitting
strategies, but its announcement rules reduce the conclusive-event rate.
}

\subsection{BBM92 protocol}

{
The BBM92 protocol, proposed by Bennett, Brassard, and Mermin in 1992, is
an entanglement-based QKD protocol \cite{bennett1992quantum}. Unlike BB84
and SARG04, a central source generates entangled photon pairs and
distributes one photon to Alice and the other to Bob, as shown in
Fig.~\ref{figure system model}.
}
{
The key is established from correlations between the two measurement
records. The polarization state belongs to the two-qubit Hilbert space
spanned by
\(\{\ket{HH},\ket{HV},\ket{VH},\ket{VV}\}\). In this work, the source is
assumed to generate the maximally entangled Bell state
}
\begin{equation}
\ket{\Phi^{+}} = \tfrac{1}{\sqrt{2}} \left( \ket{HH} + \ket{VV} \right).
\end{equation}
When both users perform their measurements in the same basis, their results are perfectly correlated. After a sufficiently large number of measurements in the $H/V$ and $D/A$ bases, Alice and Bob exchange, through an authenticated classical channel, the information regarding their chosen bases. Only the results obtained with identical bases are retained, ensuring reliable correlations. Finally, a subset of these results is publicly revealed and compared in order to estimate the QBER, defined as:  
\begin{equation}
\label{QBER_BBM92_simulé}
\text{QBER}_{\text{BBM92}} = \frac{\text{Number of erroneous correlated bits}}{\text{Total number of correlated bits}}.
\end{equation}
{
The \(11\%\) BBM92 threshold adopted from \cite{waks2002security} is used
only as a QBER comparison point, not as a stand-alone operational or
composable-security bound.
}

\section{Impact of the underwater channel on quantum transmission}
\label{Impact of the underwater channel on quantum transmission}
In an underwater environment, quantum transmission is subject to physical constraints related to the optical properties of water. Unlike free-space or atmospheric communications, an underwater optical channel undergoes significant losses that compromise the fidelity of quantum states and limit the performance of QKD protocols.
In our study, we consider the use of a collimated laser as the emission source, producing a narrow and stable beam. 
{
Therefore, we do not introduce an additional independent geometric-loss term in the link budget. However, the finite beam divergence is still retained through the effective attenuation factor used below, following the modified Beer-Lambert model.
}
The overall transmission efficiencies of the reception systems Alice and Bob can therefore be modeled as \cite{fung2006performance}
\begin{equation}
\eta_A = \eta_{\text{Alice}} \cdot A(x,\lambda),
\label{eq:etaA}
\end{equation}
\begin{equation}
\eta_B = \eta_{\text{Bob}} \cdot A(L-x,\lambda),
\label{eq:etaB}
\end{equation}
where $\eta_{\text{Alice}}$ and $\eta_{\text{Bob}}$ denote the detection efficiencies (including the quantum efficiency of the detectors), and 
{
\(A(L,\lambda)\) is the dimensionless underwater channel attenuation factor, or channel transmittance, over a propagation distance \(L\) at wavelength \(\lambda\)
}
(see subsection \ref{Atténuation sous marin}).

Three main phenomena influence the quality of transmission: attenuation, ambient irradiance, and photonic noise associated with the detectors. These effects have a direct impact on the performance of QKD protocols, particularly on the QBER, and consequently on the overall security of the transmission.
{
Accordingly, five illumination scenarios with distinct ambient-background
conditions are considered in this study and explicitly defined in Section III-C.
}
\subsection{Underwater attenuation}
\label{Atténuation sous marin}
In an underwater channel, the attenuation of the optical signal is mainly caused by two physical phenomena: absorption, resulting from the interaction of photons with water molecules that convert their energy into heat or chemical reactions, and scattering, corresponding to the deviation of photons from their initial trajectory due to suspended particles in the medium.  
These effects are characterized by the absorption coefficient $a(\lambda)$ and the scattering coefficient $b(\lambda)$, both depending on the wavelength $\lambda$, and vary depending on the type of water \mbox{(clear, coastal, or turbid) \cite{mobley1994light}.}
{
The corresponding beam attenuation, or extinction, coefficient is denoted by
\[
\alpha(\lambda)=a(\lambda)+b(\lambda),
\]
whereas \(A(L,\lambda)\) denotes the resulting dimensionless attenuation factor after propagation. Thus, \(\alpha(\lambda)\) and \(A(L,\lambda)\) represent distinct quantities.
}
{
In this study, the channel transmittance is described by the modified
Beer--Lambert model adopted in \cite{elamassie2018performance}:
\begin{equation}
A(L,\lambda)
=
\exp\left[
-\alpha(\lambda)L
\left(
\frac{d_1}{\theta L}
\right)^T
\right].
\label{eq:modified_Beer_Lambert}
\end{equation}
Here, \(\alpha(\lambda)\) is the extinction coefficient,
\(d_1\) is the transmitter-aperture diameter, and \(\theta\) is the beam
divergence angle. In the numerical calculations,
\(\theta=6^\circ\) is converted to radians, so that
\(d_1/(\theta L)\) is dimensionless.

The exponent \(T\) is an empirical correction parameter of the adopted
finite-aperture beam model. In the benchmark configurations considered
here, its value is associated with the joint aperture choice
\(d_1=d_2\), as reported in Table~\ref{Table1}, and not with the water
type. The water dependence enters separately through
\(\alpha(\lambda)\).

Equation~\eqref{eq:modified_Beer_Lambert} already incorporates the
effective geometric correction associated with beam divergence and finite
aperture. No additional independent geometric-loss factor is therefore
introduced. The value \(\theta=6^\circ\) is retained as a numerical
benchmark assumption and is not presented as a universal divergence for
underwater quantum transmitters.
}

\subsection{Photonic noise sources}
\label{Sources_photoniques_de_bruit}
{
Two independent noise contributions are considered: detector dark counts
and ambient background photons. The dark-count rate of one detector is
denoted by \(I_{\rm dc}\), in \(\mathrm{s^{-1}}\).
For a bit interval of duration \(\Delta t\), the mean number of dark counts
generated by the \(N_{\rm det}\) active detectors is

\begin{equation}
\overline{n}_{\rm dc}
=
N_{\rm det} I_{\rm dc}\Delta t .
\label{eq:mean_dark_counts}
\end{equation}
{
For BB84 and SARG04, \(N_{\rm det}=4\) active detectors are included at
Bob's receiver. For BBM92, the background probabilities \(y_A\) and \(y_B\)
are evaluated separately with \(N_{\rm det}=2\) detectors at each receiver;
the complete Alice--Bob setup therefore also contains four detectors.
}

The receiver is characterized by a circular entrance pupil of area
\begin{equation}
S_{\rm rx}
=
\frac{\pi d_2^2}{4},
\label{eq:receiver_area}
\end{equation}
where \(d_2\) is the pupil diameter. 
Let \(\delta\) denote the full angular field of view of the receiver.
In the horizontal-link geometry considered here, the receiver optical axis
is perpendicular to the downward vertical direction. Because the assumed
ambient radiance is restricted to the downward hemisphere, only the overlap
between this hemisphere and the receiver field of view (FOV) contributes to the
detected background.

For a planar entrance pupil and a receiver axis oriented horizontally, the
effective projected angular acceptance is
\begin{align}
\Omega_{\rm eff}(\delta)
&=
\int_{\mathrm{FOV}\cap\mathrm{downward\ hemisphere}}
\cos\theta\,d\Omega,
\nonumber\\
&=
\frac{\pi}{2}
\sin^2\left(\frac{\delta}{2}\right),
\label{eq:projected_solid_angle}
\end{align}
where \(\theta\) is measured from the receiver optical axis. For the
conservative full-angle assumption \(\delta=180^{\circ}\), this gives
\(\Omega_{\rm eff}=\pi/2~\mathrm{sr}\). The factor \(1/2\) results from the
overlap between the receiver-facing hemisphere and the downward hemisphere
supporting the assumed ambient radiance.
{
The full field of view \(\delta=180^\circ\) is retained as a conservative
high-background benchmark and is not intended to represent an optimized
underwater-QKD receiver.
}

{
Let \(s\in\{1,\ldots,5\}\) denote the illumination-scenario index.
For each scenario \(s\), the mean number of ambient photons incident on
the receiver during the background-acceptance window \(\Delta t'\) is,
under the narrowband approximation,
}
\begin{equation}
\overline{n}_{\rm bg}^{(s)}
=
\frac{
L_{\lambda,\downarrow}^{(s)}(z,\lambda)\,
S_{\rm rx}\,
\Omega_{\rm eff}(\delta)\,
\Delta\lambda\,
\Delta t'\,
\lambda
}{
h_p c_{\rm light}
},
\label{eq:incident_background_photons}
\end{equation}
where $h_p$ is the Planck's constant, $c_{\rm light}$ the speed of light in vaccum, \(\lambda\) the operating wavelength and \(\Delta\lambda\)
the optical-filter bandwidth. Note that the wavelength unit used for
\(\Delta\lambda\) must be consistent with that used in the definition of
\(L_{\lambda}^{(s)}\).

The corresponding mean number of detected background photons is
\begin{equation}
n_B^{(s)}
=
\eta_{\rm APD}\,
\overline{n}_{\rm bg}^{(s)},
\label{eq:detected_background_photons}
\end{equation}
where \(\eta_{\rm APD}\) is the detector quantum efficiency.

Assuming statistically independent Poissonian dark-count and ambient-photon
processes, the probability that at least one noise click occurs during a
detection window is
\begin{equation}
y_0^{(s)}
=
1-
\exp\left[
-\overline{n}_{\rm dc}
-n_B^{(s)}
\right].
\label{eq:noise_yield}
\end{equation}
In the low-count regime,
\begin{equation}
y_0^{(s)}
\simeq
N_{\rm det}I_{\rm dc}\Delta t
+
n_B^{(s)}.
\label{eq:noise_yield_low_count}
\end{equation}
{
Eq. ~\eqref{eq:noise_yield_low_count} is reported only as a low-count
interpretation. All revised numerical results use the exact Poisson
expression in Eq.~\eqref{eq:noise_yield}. Dark counts are integrated over
the bit period \(\Delta t=40~\mathrm{ns}\), whereas ambient photons are
integrated over the receiver gate \(\Delta t'=200~\mathrm{ps}\).
}

For notational simplicity, the scenario index \(s\) is omitted from
\(n_B^{(s)}\) and \(y_0^{(s)}\) in the remainder of the manuscript.

}

\subsection{
\texorpdfstring{
{Underwater Ambient Optical Background}
}{
Underwater Ambient Optical Background
}
}
\label{subsec:ambient_background}

{
To evaluate the resilience of the QKD protocols across distinct environmental
lighting conditions, we adopt five representative ambient-illumination scenarios
from the literature \cite{paglierani2023primer}. The corresponding surface-irradiance values,
\begin{equation}
E_{\mathrm{lit}}^{(s)}
\in
\left\{
10^{-3},\,10,\,50,\,125,\,500
\right\}
~\mathrm{W\,m^{-2}},
\label{eq:surface_irradiance_scenarios}
\end{equation}
are broadband visible benchmarks. They are not spectral
radiances and are not inserted directly into the narrowband photon-count
formula.}

{
The benchmark values listed in Eq.~\eqref{eq:surface_irradiance_scenarios}
represent visible-band-integrated surface irradiances from the literature. 
In a dimensionally consistent radiometric chain, however, broadband irradiances
cannot be inserted directly into a narrowband photon-count expression; they must
first be mapped to spectral irradiance, directional radiance, and ultimately 
to a detected photon flux within the receiver acceptance window.

To preserve direct comparability with existing numerical benchmarks while maintaining
radiometric rigor, we establish a calibration at equal modeled background-photon
number. Let \(n_{B,\mathrm{ref}}\) denote the background-photon count obtained under
the simplified broadband convention, and \(n_{B,\mathrm{corr}}\) the value yielded
by the spectrally and angularly resolved radiometric chain adopted in this work.
An effective broadband irradiance \(E_{\mathrm{vis,eq}}^{(s)}\) is then defined such that
\begin{equation}
n_{B,\mathrm{corr}}\!\left(E_{\mathrm{vis,eq}}^{(s)}\right)
=
n_{B,\mathrm{ref}}\!\left(E_{\mathrm{lit}}^{(s)}\right).
\label{eq:background_equivalence}
\end{equation}
Because both background-photon expressions scale linearly with incident irradiance,
this equivalence reduces to a linear scaling
\begin{equation}
E_{\mathrm{vis,eq}}^{(s)}
=
F_{\mathrm{eq}}E_{\mathrm{lit}}^{(s)},
\qquad
\text{with}\quad
F_{\mathrm{eq}}
=
\frac{n_{B,\mathrm{ref}}(1)}
     {n_{B,\mathrm{corr}}(1)}
\simeq
1.0053 \times 10^{-5}.
\label{eq:equivalent_irradiance}
\end{equation}
The calibration factor \(F_{\mathrm{eq}}\) is a model-specific parameter determined
by the chosen optical filter bandwidth, receiver solid-angle acceptance, and unit
convention, rather than a universal radiometric conversion constant.

Applying this calibration to the five reference scenarios of
Eq.~\eqref{eq:surface_irradiance_scenarios} yields the following effective surface
irradiances for our numerical evaluation:
\[
E_{\mathrm{vis,eq}}^{(s)}
\in
\left\{
1.0053\times10^{-8},\,
1.0053\times10^{-4},\,
5.0265\times10^{-4},\,
1.2566\times10^{-3},\,
5.0265\times10^{-3}
\right\}
~\mathrm{W\,m^{-2}}.
\]
Consequently, our comparative analysis is conducted at equal modeled background-photon
number rather than equal physical surface irradiance. The scaled values above serve
strictly as calibrated numerical inputs and should not be interpreted as direct
in-situ measurements of underwater solar illumination.

To model the spectral distribution of the ambient light, a uniform profile is
assumed over the visible range \(\Lambda_{\mathrm{vis}}=[380,780]~\mathrm{nm}\):
\begin{equation}
\varphi(\lambda)
=
\frac{1}{400~\mathrm{nm}},
\qquad
\int_{\Lambda_{\mathrm{vis}}}
\varphi(\lambda)\,d\lambda
=
1.
\label{eq:normalized_background_spectrum}
\end{equation}
The corresponding surface spectral irradiance, expressed in
\(\mathrm{W\,m^{-2}\,nm^{-1}}\), is given by
\begin{equation}
E_{\lambda,0}^{(s)}(\lambda)
=
E_{\mathrm{vis,eq}}^{(s)}
\varphi(\lambda).
\label{eq:surface_spectral_irradiance}
\end{equation}

As ambient light penetrates the water column, its vertical attenuation with
receiver depth \(z\) is modeled as
\begin{equation}
E_{\lambda}^{(s)}(z,\lambda)
=
E_{\lambda,0}^{(s)}(\lambda)
\exp\left(-K_{\infty}z\right),
\label{equa_irradiance}
\end{equation}
where the vertical depth \(z\) is strictly distinct from the horizontal
source--receiver propagation distance \(L\).
{
The diffuse attenuation coefficient \(K_{\infty}=0.08~\mathrm{m^{-1}}\) is
retained as a common benchmark parameter for the vertical decay of the
downwelling ambient field. Importantly, \(K_{\infty}\) is distinct from the
water-type-dependent horizontal extinction coefficient \(\alpha(\lambda)\).
Adopting a fixed \(K_{\infty}\) allows us to isolate the horizontal-link
comparison across different water types without implying identical diffuse-light
attenuation in all natural waters.

Finally, approximating the downwelling underwater field as Lambertian over the
downward hemisphere, the resulting spectral radiance, expressed in
\(\mathrm{W\,m^{-2}\,sr^{-1}\,nm^{-1}}\), is
\begin{equation}
L_{\lambda,\downarrow}^{(s)}(z,\lambda)
=
\frac{
E_{\lambda}^{(s)}(z,\lambda)
}{\pi}.
\label{eq:underwater_spectral_radiance}
\end{equation}
The angular overlap between this downwelling radiance and the horizontal
receiver field of view is accounted for separately through the effective
solid angle \(\Omega_{\rm eff}\) defined in Eq.~\eqref{eq:projected_solid_angle}.
}}

\section{
\texorpdfstring{
{Performance Metrics for the UQKD System}
}{
Performance Metrics for the UQKD System
}
}
\label{Performance metrix for UQKD Communication system}
This section presents the performance of the BB84, SARG04, and BBM92 protocols in terms of quantum gain and QBER, considering a non-turbulent underwater channel.
\subsection{Quantum gain}
Unlike idealized analytical models in which every pulse contains exactly one photon, experimental QKD transmitters use attenuated laser light. In quantum optics, these pulses are described as weak coherent states with random phase, the number of photons $N$ per pulse following a Poisson distribution with mean $\mu$. Assuming statistically independent Poissonian dark-count and ambient-photon processes, the total noise-click probability associated with one transmitted bit is
\begin{equation}
\label{Poissan}
\mathbb{P}(\text{N} = i) = \frac{\mu^i}{i!} e^{-\mu}, \quad i \in \mathbb{N}.
\end{equation}
Let $A_i$ denote the event “the pulse contains exactly $i$ photons”.
Applying the law of total probability, the quantum gain $Q_\mu$ is defined as the sum, for all photon numbers $i$, of the product of the probability that Alice emits a state with i photons, as described in \eqref{Poissan}, and the conditional probability that this state with $i$ photons (including photon noise) will result in a detection event on Bob's side, say \mbox{$Y_i \triangleq  \mathbb{P}(\text{click} \mid A_i)$.} Formally, it is expressed as \cite{ma2005practical}
\begin{align}
\label{gain_quantique_global}
Q_\mu &:= \mathbb{P}(\text{at least one detected click}), \notag \\
&= \sum_{i=0}^{\infty} \mathbb{P}(\text{click} \mid A_i) \cdot \mathbb{P}(A_i)
= \sum_{i=0}^{\infty} Y_i \cdot \frac{\mu^i}{i!} e^{-\mu}.
\end{align}

Now, to further characterise the quantum gain achievable by the system, it is necessary to specify the form of $Y_i$ as a function of the physical parameters of the system. If the pulse contains at least one photon ($i \geq 1$), the probability that none will be detected after transmission through a channel with efficiency $\eta_X$ is $(1-\eta_X)^i$, where $X \in \{A,B\}$, $A$ and $B$ representing Alice and Bob, respectively. Thus, the probability that at least one photon is detected is $1-(1-\eta_X)^i$. In this context, two events may lead to a detector click:
\begin{itemize}
\item a true detection, corresponding to the reception of at least one photon from the signal, with a probability \mbox{$1-(1-\eta_X)^i$;}
\item a false detection, independent of the signal, caused by background noise or intrinsic effects of the detector, and thus modelled by the average probability $y_0$ per pulse (see Eq.~\eqref{eq:noise_yield}).\\
\end{itemize}

\subsubsection{Quantum gain for BB84}
\label{Gain quantique BB84}
For the BB84 protocol, the total probability of detecting a pulse containing $i$ photons can be expressed as the sum of the two independent events presented above. Since $y_0$ is very small, the case where true detection and false detection occur simultaneously can be neglected.
For $i \geq 1$ and $\eta_A=1$, the total detection probability can therefore be approximated as  
\begin{align}
Y_{i}^{\text{(BB84)}} &= \mathbb{P}(\text{click} \mid A_i), \notag\\
&= \mathbb{P}(\text{true detection} \mid A_i) + \mathbb{P}(\text{false detection} \mid A_i), \notag \\
&\approx \left[1 - (1 - \eta_B)^i\right] + y_0.
\end{align}
Note that when no photon is emitted ($i = 0$), this probability reduces to $Y_{0}^{\text{(BB84)}} = y_0.$
By inserting this expression into  \eqref{gain_quantique_global}, and taking into account the Poissonian statistics of weak coherent pulses, one obtains:
\begin{equation}
\label{equa gain quantique BB84}
    Q_{\mu, \text{BB84}} = y_0 + 1 - e^{-\eta_B \mu}.
\end{equation}
This compact relation highlights the two independent mechanisms leading to a detector click. The proof is provided in Appendix-A.\\

\subsubsection{Quantum gain for SARG04}
In the SARG04 \mbox{protocol \cite{ali2012practical},} a key bit is only generated when a detector click can be interpreted as a conclusive event. Bob must then deduce with certainty the transmitted bit from his measurement basis, the result obtained, and the set of non-orthogonal states announced by Alice. {Consequently, only a fraction of the detection events contributes to the probability of a retained conclusive event.} Applying the total probability law to the number of photons \mbox{$\text{N} \sim \text{Poisson}(\mu)$,} we obtain
\begin{align}
Q_{\mu, \text{SARG04}} &= \sum_{i=0}^{\infty} \mathbb{P}(\text{conclusive click} \mid A_i)\cdot\mathbb{P}(A_i),\notag \\
&= \sum_{i=0}^{\infty} Y_i^{(\text{SARG04})}\cdot\frac{\mu^i}{i!} e^{-\mu},
\end{align}
where $Y_i^{(\text{SARG04})}$ denotes the probability of obtaining a conclusive click given that the pulse contains exactly $i$ photons.

{
\textbf{Case of no detected signal photon.}
This event includes both an empty emitted pulse and a non-empty pulse for
which all signal photons are lost during propagation. For a Poisson source
with mean photon number \(\mu\), the probability that no signal photon is
detected is \(e^{-\eta_B\mu}\). Any click occurring in this case originates
from noise with probability \(y_0\). 
Following the SARG04 announcement and conclusive-outcome rules
\cite{scarani2004quantum,fung2006performance}, we assume that a
noise-induced click is uniformly distributed over Bob's two measurement bases
and the two detector outcomes within each basis. For any fixed announced
pair and transmitted state, the four basis--outcome combinations therefore
contain one correct conclusive outcome, one erroneous conclusive outcome,
and two inconclusive outcomes. Consequently, half of the noise clicks
are conclusive, among which half are erroneous (i.e., one quarter of all noise clicks).
The corresponding contribution to the quantum gain is
}
\begin{equation}
\label{gain quan sarg04 -1-}
{
Q_{\mu,\mathrm{SARG04}}^{(\mathrm{noise})}
=
\frac{1}{2}y_0 e^{-\eta_B\mu}.
}
\end{equation}

{
\textbf{Case of at least one detected signal photon.}
For a pulse containing at least one photon, the probability of a true
detection is \(1-(1-\eta_B)^i\). Among these detections, \(1/4\)
corresponds to conclusive correct clicks and \(e_{\mathrm{det}}/2\) to
conclusive erroneous clicks. The total probability of a conclusive click
is therefore \(1/4+e_{\mathrm{det}}/2\). We then obtain
}
\begin{equation}
Y_i^{(\mathrm{SARG04})} = \left[1 - (1-\eta_B)^i\right] \left(\frac{1}{4} + \frac{e_{\mathrm{det}}}{2}\right),
\end{equation}
and 
\begin{equation}
\label{gain quan sarg04 -2-}
\begin{split}
Q_{\mu,\mathrm{SARG04}}^{(1+)} 
&= \sum_{i=1}^{\infty} Y_i^{(\mathrm{SARG04})} \frac{\mu^i}{i!} e^{-\mu} \\
&= \left( \frac{1}{4} + \frac{e_{\mathrm{det}}}{2} \right) \sum_{i=1}^{\infty} \left[1 - (1 - \eta_B)^i\right] \frac{\mu^i}{i!} e^{-\mu} \\
&= \left( \frac{1}{4} + \frac{e_{\mathrm{det}}}{2} \right) \left(1 - e^{-\eta_B \mu} \right).
\end{split}
\end{equation}
By combining both expressions \eqref{gain quan sarg04 -1-} and \eqref{gain quan sarg04 -2-}, the overall quantum gain for the SARG04 protocol is expressed as
\begin{equation}
\label{gain quantique SARG04 global}
{
Q_{\mu,\mathrm{SARG04}}
=
\underbrace{\frac{1}{2}y_0 e^{-\eta_B\mu}}_{\text{conclusive noise clicks}}
+
\underbrace{\left(\frac{1}{4}+\frac{e_{\mathrm{det}}}{2}\right)
\left(1-e^{-\eta_B\mu}\right)}_{\text{conclusive signal clicks}}.
}
\end{equation}

\subsubsection{Quantum gain for BBM92}
\label{Quantum gain for BBM92}

{
Unlike BB84 and SARG04, which are modeled using weak coherent pulses,
BBM92 relies on a source of entangled photon pairs
\cite{ma2007quantum}. We denote by \(p_{\mathrm{pair}}\) the probability
that one entangled pair is generated within an emission window. The
present analysis adopts a single-pair approximation and neglects
multipair emissions. The value \(p_{\mathrm{pair}}=1\) is used in the
numerical results as an idealized deterministic-pair benchmark and is
not intended to represent the statistics of a practical SPDC source.

Let \(y_A\) and \(y_B\) denote the probabilities that at least one
noise-induced click occurs at Alice's and Bob's receivers, respectively,
during one detection window. They are evaluated from
Eq.~\eqref{eq:noise_yield} using the detector configuration of each
receiver. For identical receivers and identical background conditions,
\(y_A=y_B=y_0\).

Conditioned on the generation of one pair, the probability that receiver
\(X\in\{A,B\}\) records a click is
\begin{equation}
c_X
=
\eta_X+(1-\eta_X)y_X.
\label{eq:BBM92_receiver_click}
\end{equation}
The first term corresponds to detection of the signal photon, whereas the
second corresponds to a noise click when the signal photon is not detected.
This expression avoids counting a simultaneous signal and noise arrival as
two independent receiver clicks.

The exact coincidence probability used in the numerical calculations is
therefore
\begin{align}
Q_{\mathrm{BBM92}}
&=
\mathbb{P}_{\mathrm{coinc}}
\notag\\
&=
p_{\mathrm{pair}}\,c_Ac_B
+
(1-p_{\mathrm{pair}})\,y_Ay_B
\notag\\
&=
p_{\mathrm{pair}}
\left[\eta_A+(1-\eta_A)y_A\right]
\left[\eta_B+(1-\eta_B)y_B\right]
+
(1-p_{\mathrm{pair}})y_Ay_B.
\label{Pcoin}
\end{align}

A true coincidence occurs when the two photons originating from the same
generated pair are both detected:
\begin{equation}
\mathbb{P}_{\mathrm{true}}
=
p_{\mathrm{pair}}\eta_A\eta_B.
\label{eq:BBM92_true_coincidence}
\end{equation}
All remaining coincidence events are classified as false coincidences:
\begin{align}
\mathbb{P}_{\mathrm{false}}
&=
\mathbb{P}_{\mathrm{coinc}}
-
\mathbb{P}_{\mathrm{true}}
\notag\\
&=
p_{\mathrm{pair}}
\Big[
\eta_A(1-\eta_B)y_B
+
\eta_B(1-\eta_A)y_A
+
(1-\eta_A)(1-\eta_B)y_Ay_B
\Big]
\notag\\
&\quad
+
(1-p_{\mathrm{pair}})y_Ay_B.
\label{eq:BBM92_false_coincidence}
\end{align}
Consequently,
\begin{equation}
\mathbb{P}_{\mathrm{coinc}}
=
\mathbb{P}_{\mathrm{true}}
+
\mathbb{P}_{\mathrm{false}}.
\label{eq:BBM92_coincidence_partition}
\end{equation}

This coincidence model is distinct from the Poisson weak-coherent-pulse
description used for BB84 and SARG04. In particular, the mean coherent-state
photon number \(\mu\) does not enter the BBM92 expressions. The former
first-order approximation
\(p_{\mathrm{pair}}y_0(\eta_A+\eta_B)+y_0^2\)
is not used to generate the revised numerical results.
}

\subsection{QBER}
\label{QBER Analysis}
{
Following the standard gain/error-gain formulation used in practical QKD
analyses \cite{ma2005practical,ali2012practical}, the QBER is defined as
the ratio between the erroneous-event gain and the total quantum gain:
}
\begin{align}
    \text{QBER}
    =
    \frac{E_\mu}{Q_\mu},
\end{align}
{
where \(Q_\mu\) denotes the total quantum gain and \(E_\mu\) denotes the
contribution of erroneous retained events to that gain.

A high QBER value may indicate either excessive noise in the system or an active eavesdropping attempt. In the following, we derive the analytical expressions of the QBER for each of the considered protocols (BB84, SARG04, and BBM92), based on the previously established expressions of the quantum gain.}\\

\subsubsection{QBER for BB84}  
In the BB84 protocol, errors mainly originate from two sources.  
The first corresponds to errors in detecting signal photons due to sensor imperfections, characterised by an error probability \( e_{\text{det}} \).  
The second source is related to false clicks, i.e., detection events triggered by noise photons of the photodetectors. Unlike signal photons, these clicks occur randomly and therefore carry no information about the prepared polarization state. Since each measurement basis (\( H/V \) or \( D/A \)) involves two detectors, a spurious click has a probability of $50\%$ of producing a correct bit and $50\%$ of introducing an error. Consequently, the associated error probability is \( e_0 = \tfrac{1}{2} \).

The total errors in the BB84 protocol can therefore be expressed as
\begin{equation}  
\label{Erreur BB84}  
    E_{\mu, \text{BB84}} = \frac{1}{2} y_0 + e_{\text{det}} \left(1 - e^{-\eta_B \mu} \right).
\end{equation} 

{
Substituting the BB84 quantum gain given in
\eqref{equa gain quantique BB84}, the QBER is written explicitly as
\begin{align}
\label{QBER BB84}
\text{QBER}_{\text{BB84}}
=
\dfrac{
e_0 y_0
+
e_{\text{det}}\left(1-e^{-\eta_B\mu}\right)
}{
y_0+1-e^{-\eta_B\mu}
}.
\end{align}
}

\subsubsection{QBER for SARG04}
In the SARG04 protocol, only conclusive detection events are used for key generation. This specificity modifies both the expression of the gain and the error structure compared to the BB84 protocol. The QBER is therefore defined as the ratio between the total number of errors from conclusive detections and the overall gain of the protocol. Two types of errors are considered. The first originates from false clicks: as in BB84, they occur randomly, but only a fraction of them is retained as conclusive; 
{ as derived in the preceding SARG04 gain analysis from the protocol's
conclusive-outcome rules \cite{scarani2004quantum,fung2006performance},} one quarter of these clicks are both retained and erroneous.
The second source is related to the detection of signal photons: only conclusive detections are taken into account, and half of the errors associated with these events are effectively preserved. The total number of errors is obtained by combining these two terms. 
{
Substituting the SARG04 quantum gain given in
\eqref{gain quantique SARG04 global}, the QBER is written explicitly as
\begin{align}
\label{QBER SARG04}
\text{QBER}_{\text{SARG04}}
=
\dfrac{
    \tfrac{1}{4} y_{0} e^{-\eta_B \mu}
    +
    \tfrac{e_{\text{det}}}{2}
    \left(1-e^{-\eta_B\mu}\right)
}{
    \tfrac{1}{2} y_{0}e^{-\eta_B\mu}
    +
    \left(
        \tfrac{1}{4}
        +
        \tfrac{e_{\text{det}}}{2}
    \right)
    \left(1-e^{-\eta_B\mu}\right)
}.
\end{align}
}

\subsubsection{QBER for BBM92}

{
In the BBM92 protocol, the QBER is evaluated from the coincidence
detections recorded by Alice and Bob
\cite{muskan2023analysing}. Two classes of events contribute to the
error probability. True coincidences originate from the two photons of
the same generated pair and have the conditional signal-error probability denoted as
\(e_{\mathrm{sig}}\).
False coincidences contain at least
one noise-induced detection and are assumed to produce a random bit, with
an error probability equal to \(1/2\).

Using the coincidence partition introduced in
Eq.~\eqref{eq:BBM92_coincidence_partition}, the BBM92 QBER is
\begin{equation}
\label{QBER BBM92}
\mathrm{QBER}_{\mathrm{BBM92}}
=
\frac{
e_{\mathrm{sig}}\,
\mathbb{P}_{\mathrm{true}}
+
\tfrac{1}{2}\,
\mathbb{P}_{\mathrm{false}}
}{
\mathbb{P}_{\mathrm{coinc}}
}
=
\frac{
e_{\mathrm{sig}}\,
\mathbb{P}_{\mathrm{true}}
+
\tfrac{1}{2}\,
\mathbb{P}_{\mathrm{false}}
}{
Q_{\mathrm{BBM92}}
},
\end{equation}}

{
where we recall that
$
\mathbb{P}_{\mathrm{true}} =p_{\mathrm{pair}}\eta_A\eta_B, \;
\mathbb{P}_{\mathrm{false}} =p_{\mathrm{pair}}
\Big[
\eta_A(1-\eta_B)y_B
+
\eta_B(1-\eta_A)y_A
+
(1-\eta_A)(1-\eta_B)y_Ay_B
\Big]
+
(1-p_{\mathrm{pair}})y_Ay_B,$ and 
$Q_{\mathrm{BBM92}}=p_{\mathrm{pair}}
\left[\eta_A+(1-\eta_A)y_A\right]
\left[\eta_B+(1-\eta_B)y_B\right]
+
(1-p_{\mathrm{pair}})y_Ay_B.$

The position of the entangled-photon source affects the BBM92 QBER through
two coupled mechanisms: the balance between the arm detection efficiencies
and the position dependence of the conditional polarization error. For
identical receivers and a homogeneous water type, exchanging the two arm
lengths, \(x\leftrightarrow L-x\), leaves the complete QBER expression
unchanged. The midpoint \(x=L/2\) is therefore a stationary configuration,
but symmetry alone does not guarantee that it is the global optimum for
every background-noise level.

For the parameter ranges considered here, the numerical evaluation shows
that the midpoint provides the largest QBER-threshold distance in
Scenarios~1--4. In the background-dominated Scenario~5, the dependence on
the source position becomes very weak and a marginally larger threshold
distance may occur near an end position. The symmetry argument and the
competition between polarization degradation and accidental coincidences
are detailed in Appendix~\ref{Appendix B}.
}

\section{Evolution of the Entangled State Using Kraus Operators in an Underwater Quantum Channel}
\label{Modeling Entangled States with Kraus Operators in an Underwater Quantum Channel}
This section extends the classical underwater-channel model to a quantum framework in order to describe how environmental parameters influence the evolution of an entangled state. The underwater medium is modeled as a noisy open quantum system, where photon interactions with the environment induce decoherence and irreversible dynamics.

In this study, Alice and Bob share a pair of qubits prepared in a maximally entangled Bell state, represented by the density matrix $\rho_{AB}^{\text{in}} = \ket{\Phi^+}\bra{\Phi^+}$,
expressed in the canonical basis ${\ket{00}, \ket{01}, \ket{10}, \ket{11}}$, which forms an orthonormal basis of the bipartite Hilbert space $\mathcal{H}_{AB} = \mathcal{H}_A \otimes \mathcal{H}_B$.
In an ideal isolated system, this state evolves as 
$\rho_{AB}^{\text{in}} \rightarrow U\rho_{AB}^{\text{in}}U^{\dagger}$ with $U$ a unitary operator,
preserving both information and entanglement \cite{holevo2019quantum}. 
In contrast, when transmitted through an underwater channel, the evolution becomes non-unitary and is described by a completely positive and trace-preserving (CPTP) map represented by a set of Kraus operators $\{K_j\}$. 
The number of Kraus operators ‘$n$’ depends on the dimension ‘$d$’ of the Hilbert space of the system, such that $n \leq d^2$. 
{
For a qubit, \(d=2\), and a general quantum channel can be represented using
at most \(d^2=4\) Kraus operators
\cite{wilde2013quantum,holevo2019quantum}. In the model considered below,
photon loss is included through the arm detection efficiencies, whereas
polarization degradation is described by local depolarizing channels.
}

{
Each Kraus operator represents one contribution to the system--environment
interaction. For a channel acting on the bipartite polarization space,
\begin{equation}
\rho_{AB}^{\mathrm{out}}
=
\mathcal{E}(\rho_{AB}^{\mathrm{in}})
=
\sum_{j=0}^{n-1}K_j\rho_{AB}^{\mathrm{in}}K_j^\dagger,
\label{Kraus_oper_general}
\end{equation}
where the global Kraus operators satisfy
\[
\sum_{j=0}^{n-1}K_j^\dagger K_j=I_A\otimes I_B.
\]
If the channel acts only on Bob's photon, the corresponding global
operators are \(I_A\otimes K_j^{(B)}\). In the two-arm model used below,
they are the products \(L_i(q_A)\otimes L_j(q_B)\).
}

{
In the model adopted here, photon loss and polarization degradation are
treated as distinct physical mechanisms. Absorption and scattering outside
the receiver acceptance remove photons from the detected polarization-qubit
subspace. These losses are included only through the channel transmittances
and the corresponding detection efficiencies \(\eta_A\) and \(\eta_B\).
They are not applied a second time as an amplitude-damping map within the
two-dimensional polarization space.

Conditioned on the successful detection of both photons, the surviving
polarization state is subsequently affected by independent local
depolarizing channels, as described in
Section~\ref{depolarizing channel}. Consequently, no ordering between an
amplitude-damping channel and a depolarizing channel is required in the
revised model.
}

\subsection{
\texorpdfstring{
{Photon Loss and Conditional Polarization State}
}{
Photon Loss and Conditional Polarization State
}
}
\label{amplitude channel}

{
In polarization-encoded BBM92, absorption and scattering outside the
receiver aperture or field of view correspond primarily to photon-loss
events rather than to transitions between the logical polarization states
\(\ket{H}\) and \(\ket{V}\). A lost photon leaves the detected
polarization-qubit subspace and is formally represented by an erasure
into an orthogonal vacuum state \(\ket{\mathrm{vac}}\)
\cite{wilde2013quantum,holevo2019quantum}.

For a propagation arm \(X\in\{A,B\}\), let \(A_X\) denote the corresponding
power transmittance,
\[
A_A=A(x,\lambda),
\qquad
A_B=A(L-x,\lambda).
\]
For an input state supported on the polarization subspace, the photon-loss
channel mapping to the enlarged polarization-plus-vacuum space can be
written formally as
\begin{equation}
\mathcal{L}_X(\rho)
=
A_X\rho
+
\left(1-A_X\right)
\operatorname{Tr}(\rho)
\ket{\mathrm{vac}}\bra{\mathrm{vac}},
\qquad X\in\{A,B\},
\label{eq:photon_loss_channel}
\end{equation}
where \(\ket{\mathrm{vac}}\) is orthogonal to the logical qubit subspace.
The corresponding photon-loss probabilities are therefore
\[
p_{\mathrm{loss},A}=1-A(x,\lambda),
\qquad
p_{\mathrm{loss},B}=1-A(L-x,\lambda).
\]

In the operational model used for the QBER calculation, vacuum outcomes
are not retained as polarization measurements. Instead, photon loss is
incorporated directly and exclusively through the overall arm detection
efficiencies defined in Eqs.~\eqref{eq:etaA} and~\eqref{eq:etaB}, namely
\(\eta_A = \eta_{\mathrm{Alice}}A(x,\lambda)\) and
\(\eta_B = \eta_{\mathrm{Bob}}A(L-x,\lambda)\).
Thus, a pair contributes to a true signal coincidence only when both photons
survive propagation and are detected, an event occurring with a probability
proportional to \(\eta_A\eta_B\).

Under the polarization-independent-loss assumption adopted in this work,
pure optical attenuation does not transfer population between
\(\ket{H}\) and \(\ket{V}\). Conditioned on the successful detection of
both photons, the bipartite polarization density matrix entering the
noise model remains uncorrupted by loss:
\begin{equation}
\rho_{AB}^{\mathrm{cond}}
=
\rho_{AB}^{\mathrm{in}}
=
\ket{\Phi^+}\bra{\Phi^+}.
\label{rho_AB_conditional}
\end{equation}
Polarization degradation of this conditional state is modeled separately
by the local depolarizing channels introduced in the following subsection.

Crucially, this separation prevents underwater attenuation from being
counted twice—once in the detection efficiencies and again within the
conditional polarization state. It also establishes a clear physical
distinction between a reduction in the coincidence detection rate and a
degradation of the polarization correlations measured on surviving
photon pairs.
}

\subsection{Depolarizing Channel}
\label{depolarizing channel}
{
The depolarizing channel considered here is modeled as an isotropic
Pauli-error channel \cite{wilde2013quantum,holevo2019quantum}. 
With probability \(1-q\), the polarization state is left unchanged, whereas
with total probability \(q\), one of the three Pauli operations is applied,
each with probability \(q/3\). This process progressively reduces polarization
coherence and quantum correlations.
}

For a single propagation arm, the isotropic depolarizing map acting locally
on one subsystem of the bipartite polarization state is written as
\begin{equation}
\label{canal dep1}
\mathcal{E}_{\text{dep}}(\rho_{AB}^\text{in}) = (1 - q) \rho_{AB}^\text{in} + \frac{q}{3} \sum_{i=x,y,z} (I_A \otimes \sigma_i)\, \rho_{AB}^\text{in}\, (I_A \otimes \sigma_i),
\end{equation}
where the operators \(\sigma_i\), \(i\in\{x,y,z\}\), denote the standard Pauli matrices:
\begin{equation}
\label{pauli_matrices}
\sigma_x =
\begin{pmatrix}
0 & 1 \\ 
1 & 0
\end{pmatrix}, \quad
\sigma_y =
\begin{pmatrix}
0 & -i \\ 
i & 0
\end{pmatrix}, \quad
\sigma_z =
\begin{pmatrix}
1 & 0 \\ 
0 & -1
\end{pmatrix}.
\end{equation}

We adopt an operational representation of Eq.~\eqref{canal dep1} to describe the
evolution of the system using four Kraus operators \(\{L_i(q)\}\) defined as
\begin{equation}
\label{local_kraus_operators}
L_0(q) = \sqrt{1 - q}\, I, \quad  
L_1(q) = \sqrt{\frac{q}{3}}\, \sigma_x, \quad
L_2(q) = \sqrt{\frac{q}{3}}\, \sigma_y, \quad
L_3(q) = \sqrt{\frac{q}{3}}\, \sigma_z,
\end{equation}
with \(I\) the identity matrix. In the two-arm BBM92 configuration, where Alice's
and Bob's photons propagate through independent local depolarizing channels, the
global Kraus operators \(\{L_{ij}\}\) acting on the bipartite system are given by
the tensor products of the local operators:
\begin{equation}
L_{ij} = L_i(q_A) \otimes L_j(q_B), \quad i,j \in \{0,1,2,3\}.
\end{equation}
{
Here, the effective depolarization probabilities associated with Alice's and Bob's
propagation arms are parameterized respectively as
\begin{equation}
q_A = q(x) = 1 - e^{-\gamma_{\mathrm{dep}}x},
\qquad
q_B = q(L-x) = 1 - e^{-\gamma_{\mathrm{dep}}(L-x)},
\label{eq:qA_qB_param}
\end{equation}
where \(\gamma_{\mathrm{dep}}\) is a phenomenological distance-to-error mapping
parameter.

Applying these local channels to the conditional Bell state yields the output
polarization state
\begin{align}
\rho_{AB}^{\mathrm{out-dep}}
&=
\sum_{i,j=0}^{3}
L_{ij}\rho_{AB}^{\mathrm{cond}}L_{ij}^{\dagger}
\notag\\
&=
\frac{1}{4}
\begin{pmatrix}
1+r & 0 & 0 & 2r \\
0 & 1-r & 0 & 0 \\
0 & 0 & 1-r & 0 \\
2r & 0 & 0 & 1+r
\end{pmatrix},
\label{equa rho depo}
\end{align}
with
\begin{equation}
\kappa_A = 1 - \frac{4q_A}{3},
\qquad
\kappa_B = 1 - \frac{4q_B}{3},
\qquad
r = \kappa_A\kappa_B.
\label{eq:depolarizing_contraction}
\end{equation}
This density matrix preserves unit trace and complete positivity, and is utilized
in our numerical evaluations. The explicit 16-term derivation and completeness checks
are detailed in Appendix~\ref{appendix:BBM92_depolarization}.

Importantly, the exponential form adopted in Eq.~\eqref{eq:qA_qB_param} is an
effective one-parameter assumption rather than a microscopic physical law or a
Beer-Lambert attenuation model. In our numerical study, this mapping is applied
strictly within the weak-depolarization regime (\(q \ll 1\)), where
\(q(\ell) \simeq \gamma_{\mathrm{dep}}\ell\). No extrapolation toward \(q \to 1\)
is attempted or interpreted as physical saturation; indeed, within the isotropic
Pauli parameterization, the completely mixed state occurs at \(q = 3/4\) rather
than \(q = 1\).

Because direct measurements of the depolarization coefficient at \(\lambda = 530~\mathrm{nm}\)
are not available for the three considered water types, we treat \(\gamma_{\mathrm{dep}}\)
as an effective benchmark parameter representing increasing polarization degradation.
Specifically, we adopt the numerical values
\[
\gamma_{\mathrm{dep}}
\in
\left\{
2.4\times10^{-6},\;
3.7\times10^{-6},\;
7.5\times10^{-6}
\right\}
~\mathrm{m^{-1}}
\]
for clear, coastal, and turbid waters, respectively. These values are scaled order-of-magnitude
estimates derived from the \(810~\mathrm{nm}\) underwater experiment reported in
\cite{ji2017towards} (as detailed in Appendix~D), rather than directly measured
material constants. Consequently, all BBM92 performance predictions involving
polarization degradation should be interpreted as conditional on these effective
parameter choices.
}
\subsection{QBER analysis for BBM92 using Kraus operators}
\label{QBER Analysis avec Kraus operateurs}
{
While Eq.~\eqref{QBER BBM92} expresses the overall BBM92 QBER in terms of true
and false coincidences, this subsection derives the conditional signal-error
probability \(e_{\mathrm{sig}}\) from the depolarized bipartite state of
Eq.~\eqref{equa rho depo}.

As established above, the detection efficiencies \(\eta_A\) and \(\eta_B\)
govern the true- and false-coincidence probabilities, and consequently the
detected event rate. The Kraus map determines exclusively the conditional
polarization error of detected true coincidences. Crucially, no Kraus-derived
survival probability is multiplied into the coincidence rates, ensuring that
optical attenuation is not counted twice.

For the two retained BBM92 measurement bases, evaluating the Pauli observables
on the state in Eq.~\eqref{equa rho depo} yields
\begin{equation}
C_X
=
\mathrm{Tr}\!\left[
\rho_{AB}^{\mathrm{out-dep}}(\sigma_x\otimes\sigma_x)
\right]
= r,
\qquad
C_Z
=
\mathrm{Tr}\!\left[
\rho_{AB}^{\mathrm{out-dep}}(\sigma_z\otimes\sigma_z)
\right]
= r,
\label{eq:BBM92_correlations}
\end{equation}
whereas \(C_Y = -r\). Thus, the \(X\)- and \(Z\)-basis correlations undergo
an identical contraction within the isotropic local-depolarizing model.

For the ideal maximally entangled state \(\ket{\Phi^+}\), \(C_X = C_Z = 1\).
The corresponding basis-dependent error probabilities are given by
\begin{equation}
e_X = \frac{1-C_X}{2},
\qquad
e_Z = \frac{1-C_Z}{2},
\end{equation}
and their equal-basis average is
\begin{equation}
\label{Kraus probability both bases}
\mathbb{P}_{\mathrm{Kraus}}
=
\frac{e_X+e_Z}{2}
=
\frac{2-C_X-C_Z}{4}.
\end{equation}
Here, \(\mathbb{P}_{\mathrm{Kraus}}\) represents the channel-induced bit-error
probability averaged over the two retained bases, rather than the probability
that an arbitrary Kraus error operation occurred. Assuming statistically
independent detector bit flips characterized by error rate \(e_{\mathrm{det}}\),
the total conditional signal-error probability is
\begin{equation}
\label{eq:BBM92_signal_error}
e_{\mathrm{sig}}
=
e_{\mathrm{det}}
+ \left(1-2e_{\mathrm{det}}\right)
\mathbb{P}_{\mathrm{Kraus}}.
\end{equation}
}

\section{Results and Discussion}
\label{Results}

{
To evaluate and compare the performance of the BB84, SARG04, and BBM92 protocols,
the analytical QBER expressions derived in Sections~\ref{QBER Analysis}
and~\ref{QBER Analysis avec Kraus operateurs} are analyzed across distinct water
turbidity types and ambient-illumination scenarios. Throughout this section,
the threshold distance \(L_{\mathrm{th}}\) denotes the maximum propagation distance
at which the modeled QBER remains below the adopted protocol-dependent threshold.
This metric serves strictly as a QBER-based comparative indicator; it should not be
interpreted as a complete secret-key-security bound, as our analysis does not
incorporate finite-key effects, error-correction leakage, privacy-amplification
inefficiencies, or composable-security proofs.

The numerical evaluations are conducted using the representative optical and system
parameters summarized in Table~\ref{Table1}, which are compiled from recent
underwater QKD literature \cite{paglierani2023primer, ata2025impact}. Notably, the
value \(p_{\mathrm{pair}}=1\) adopted for BBM92 is an idealized single-pair emission
benchmark used to probe baseline coincidence dynamics, rather than an experimentally
measured source parameter.
}

\begin{table}[!t]
\centering
\caption{System and channel parameters}
\label{Table1}
\small
\setlength{\tabcolsep}{4pt}
\begin{tabular}{@{}p{0.13\textwidth}p{0.60\textwidth}p{0.19\textwidth}@{}}
\toprule
Parameter & Definition & Value \\ 
\midrule
$\lambda$        & Wavelength\cite{paglierani2023primer}                                   & 530 nm \\
\(\theta\) & Beam-divergence angle (converted to radians in calculations)
& \(6^\circ\)\\
 {$\delta$}& {Full receiver field-of-view angle}\cite{paglierani2023primer}&  {180$^\circ$} \\
$\Delta\lambda$  & Filter spectral width\cite{ata2025impact}                       & 0.2 nm \\
$\Delta t$       & Bit period\cite{ata2025impact}                                  & 40 ns \\
$\Delta t^{'}$   & Receiver gate time \cite{ata2025impact}                          & 200 ps \\
$d_1$            & Transmitter pupil diameter                   & {5, 10, 20, 30 cm} \\
$d_2$            & Receiver pupil diameter                      & {5, 10, 20, 30 cm} \\
{$\eta_{\rm APD}$} & {Quantum efficiency of the Geiger-mode APDs} \cite{paglierani2023primer} &  {0.5} \\
 {$\mu$}& {Mean photon number per weak coherent pulse (BB84 and SARG04)} &  {1} \\
{$p_{\mathrm{pair}}$} & {Entangled-pair generation probability per emission window (BBM92)} & {1} \\
$K_\infty$       & Asymptotic diffuse attenuation coefficient\cite{paglierani2023primer}    & 0.08 m$^{-1}$ \\
$I_{dc}$         & Dark current count rate \cite{paglierani2023primer}                      & 60 Hz \\
{$z$} & {Vertical receiver depth below the water surface}\cite{ata2025impact}& {80 m} \\
$e_\text{det}$   & Detection error rate \cite{paglierani2023primer}                        & 3.3\% \\
\addlinespace[4pt]$\alpha$         & Extinction coefficient\cite{mobley1994light}                     &  \\[-2pt]
                 & \qquad Clear water                            & 0.151 m$^{-1}$ \\
                 & \qquad Coastal water                          & 0.339 m$^{-1}$ \\
                 & \qquad Turbid water                          & 2.195 m$^{-1}$ \\
\addlinespace[4pt]
$T$ & Empirical correction exponent & \\[-2pt]
                 & \quad $d_1 = d_2 = 5$ cm                     & 0.13 \\
                 & \quad $d_1 = d_2 = 10$ cm                    & 0.16 \\
                 & \quad $d_1 = d_2 = 20$ cm                    & 0.21 \\
                 & \quad $d_1 = d_2 = 30$ cm                    & 0.26 \\
\addlinespace[4pt]
{$\gamma_\text{dep}$} & {Effective depolarization coefficient used in the benchmark} & \\[-2pt]
 & \qquad Clear water  & $\scalebox{0.9}{$2.4\times 10^{-6} \text{m}^{-1}$}$ \\
 & \qquad Coastal water & $\scalebox{0.9}{$3.7\times 10^{-6} \text{m}^{-1}$}$ \\
 & \qquad Turbid water & $\scalebox{0.9}{$7.5\times 10^{-6} \text{m}^{-1}$}$ \\
\bottomrule
\end{tabular}
\end{table}

{
\subsection*{Numerical Implementation Consistency Checks}
To verify the mathematical and software accuracy of the analytical models, all
theoretical QBER curves are compared against stochastic Monte Carlo estimates.
Because these simulations operate directly on the event probabilities derived in the
preceding sections, they serve as implementation-consistency checks rather than
independent physical validations of the underlying underwater propagation model;
specifically, they do not simulate ray-tracing trajectories, multiple scattering,
or turbulence.

For BB84 and SARG04, stochastic estimates are generated via binomial resampling of
the analytical QBER. For each parameter configuration and distance, \(10^{4}\)
independent repetitions of \(10^{3}\) conclusive events are simulated, and the error
ratio is averaged over all trials.

For the entanglement-based BBM92 protocol, simulations are executed at the
emission-window level. Because the coincidence probability \(\mathbb{P}_{\mathrm{coinc}}\)
decreases rapidly with propagation distance, the number of simulated windows per
repetition is selected adaptively to yield an expected sample size of approximately
\(10^{3}\) coincidence events:
\begin{equation}
N_{\mathrm{win}}
=
\max\left[
10^{4},
\left\lceil
\frac{10^{3}}{\mathbb{P}_{\mathrm{coinc}}}
\right\rceil
\right].
\label{eq:BBM92_MC_windows}
\end{equation}
In each of the 300 independent repetitions, a multinomial draw assigns the
\(N_{\mathrm{win}}\) emission windows to five mutually exclusive categories — true correct,
true erroneous, false correct, and false erroneous coincidences, as well as
non-coincidence events — with respective probabilities
\begin{equation}
\mathbb{P}_{\mathrm{true}}(1-e_{\mathrm{sig}}),
\quad
\mathbb{P}_{\mathrm{true}}e_{\mathrm{sig}},
\quad
\frac{1}{2}\mathbb{P}_{\mathrm{false}},
\quad
\frac{1}{2}\mathbb{P}_{\mathrm{false}},
\quad\text{and}\quad
1-\mathbb{P}_{\mathrm{coinc}}.
\label{eq:BBM92_MC_categories}
\end{equation}
The simulated BBM92 QBER is then evaluated from the recorded event counts as
\begin{equation}
\widehat{\mathrm{QBER}}_{\mathrm{BBM92}}
=
\frac{
N_{\mathrm{true,err}}
+
N_{\mathrm{false,err}}
}{
N_{\mathrm{true}}
+
N_{\mathrm{false}}
}.
\label{eq:BBM92_MC_QBER}
\end{equation}
In addition, the conditional polarization correlations presented in Fig.~\ref{figure 6}
are verified separately by sampling \(2\times 10^{5}\) binary Pauli-basis outcomes
\((X, Y, Z)\) per repetition.
}

\begin{figure}[!t]
    \centering
    
    \begin{minipage}{0.49\textwidth}
    
        \centering
        \includegraphics[width=\linewidth]{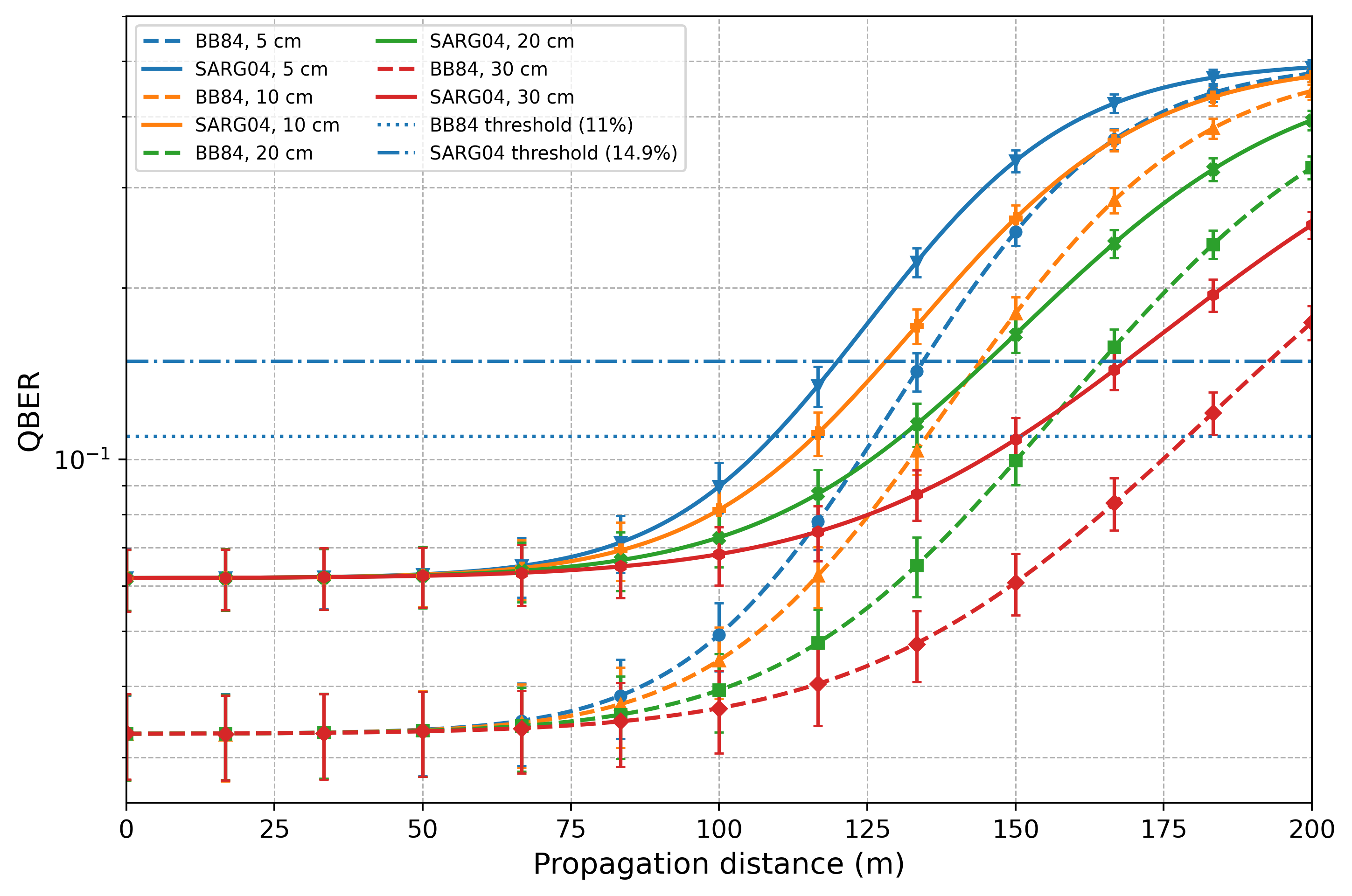}
        \\ (a)
    \end{minipage}
    \hfill
    \begin{minipage}{0.49\textwidth}
    
        \centering
        \includegraphics[width=\linewidth]{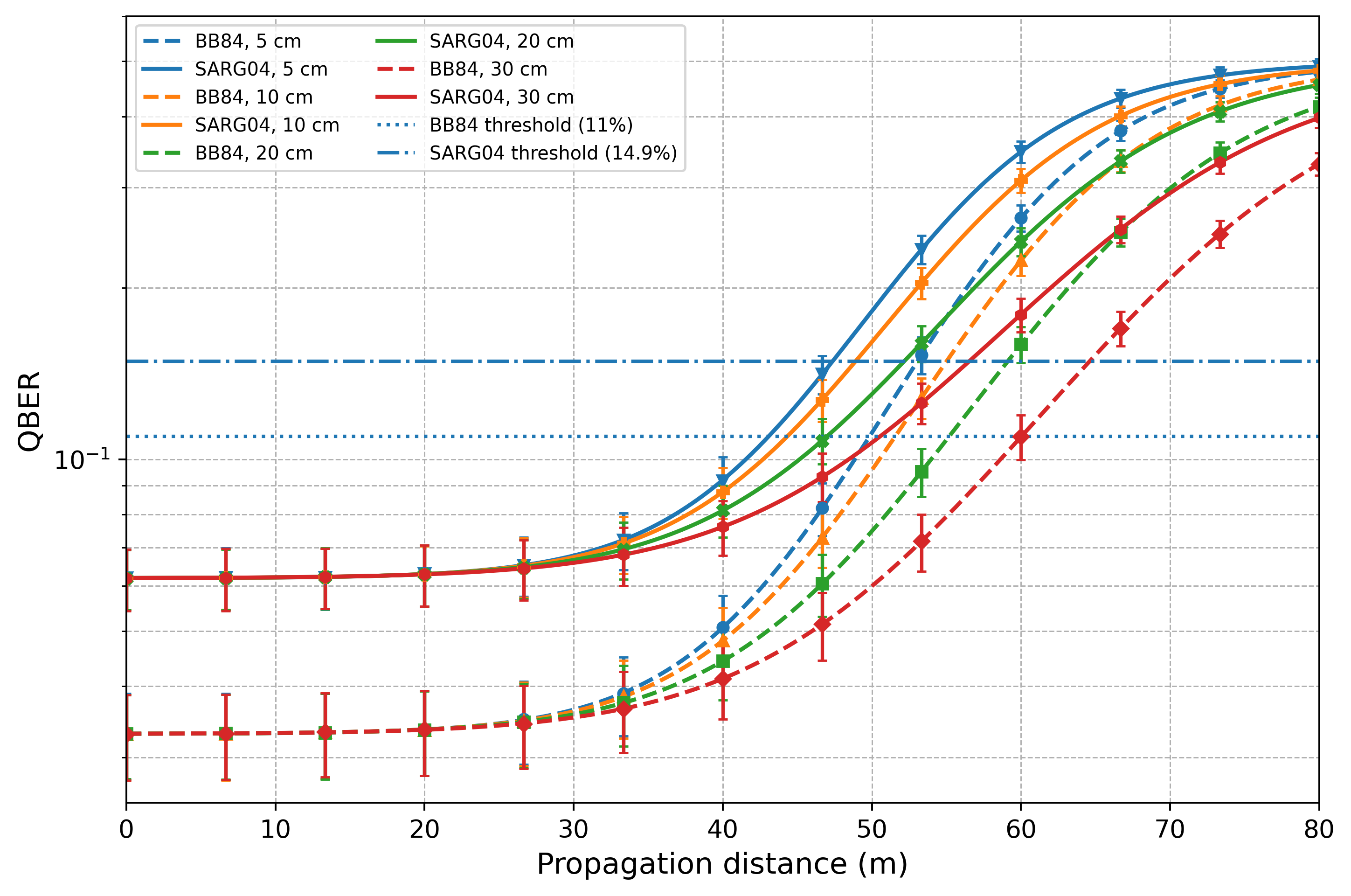}
        \\ (b)
    \end{minipage}
    \hfill
    \begin{minipage}{0.49\textwidth}
    
        \centering
        \includegraphics[width=\linewidth]{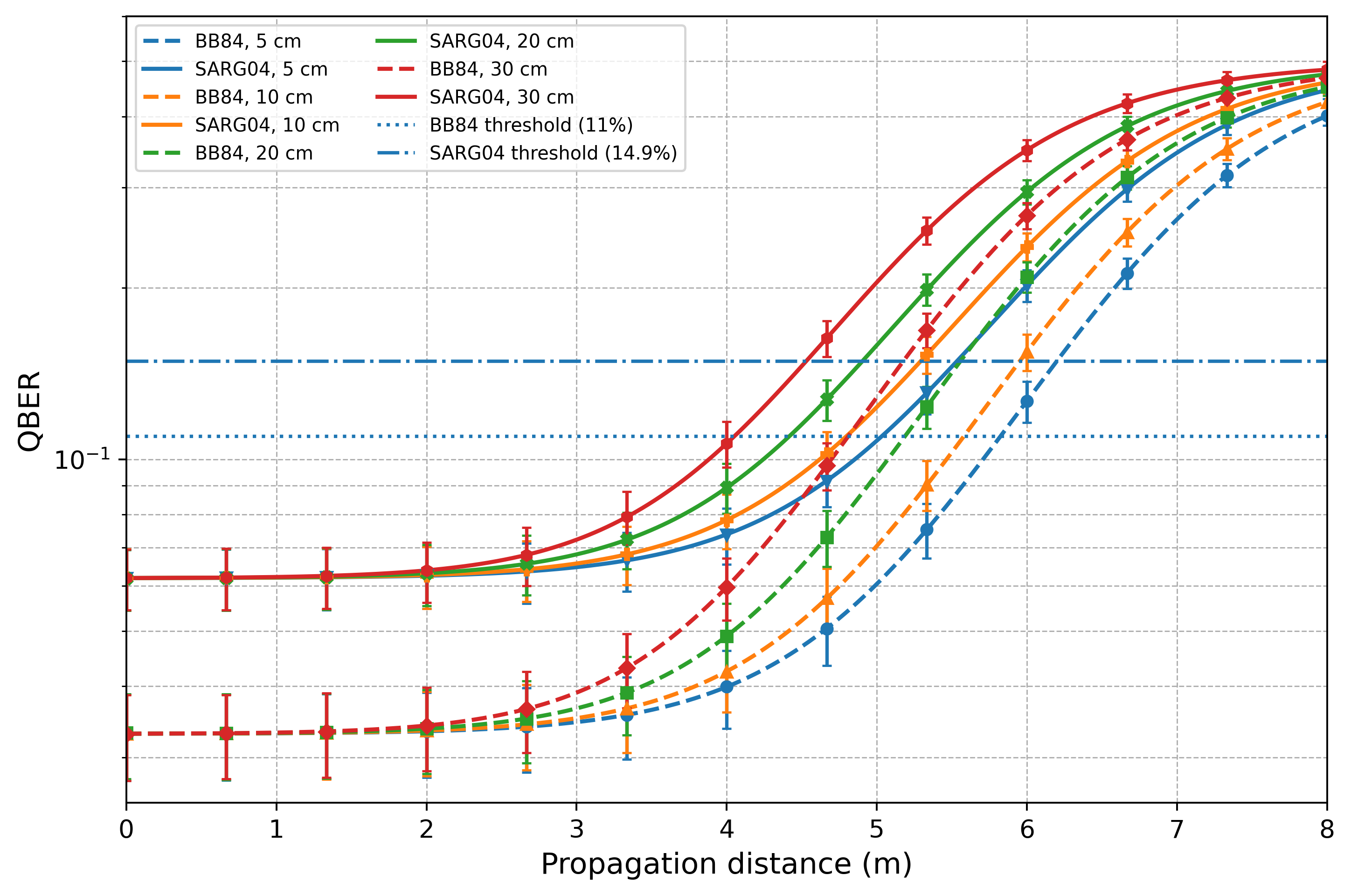}
        \\ (c)
    \end{minipage}
    
 \caption{{\small
QBER$_{\mathrm{BB84}}$ and QBER$_{\mathrm{SARG04}}$ versus propagation
distance for Scenario~1, defined by the literature broadband-irradiance
benchmark
\(E_{\mathrm{lit}}=10^{-3}\,\mathrm{W\,m^{-2}}\)
and mapped to the equivalent radiometric input through
Eq.~\eqref{eq:equivalent_irradiance}. The different curves correspond to
\(d_1=d_2=5\), 10, 20, and 30~cm in (a) clear, (b) coastal, and
(c) turbid water. Analytical curves follow
Eqs.~\eqref{QBER BB84} and~\eqref{QBER SARG04}; markers denote Monte Carlo
consistency estimates.
}}

\label{figure 1}
\end{figure}

\begin{figure}[!t]
    \centering
    
    \begin{minipage}{0.49\textwidth}
        \centering
        \includegraphics[width=\linewidth]{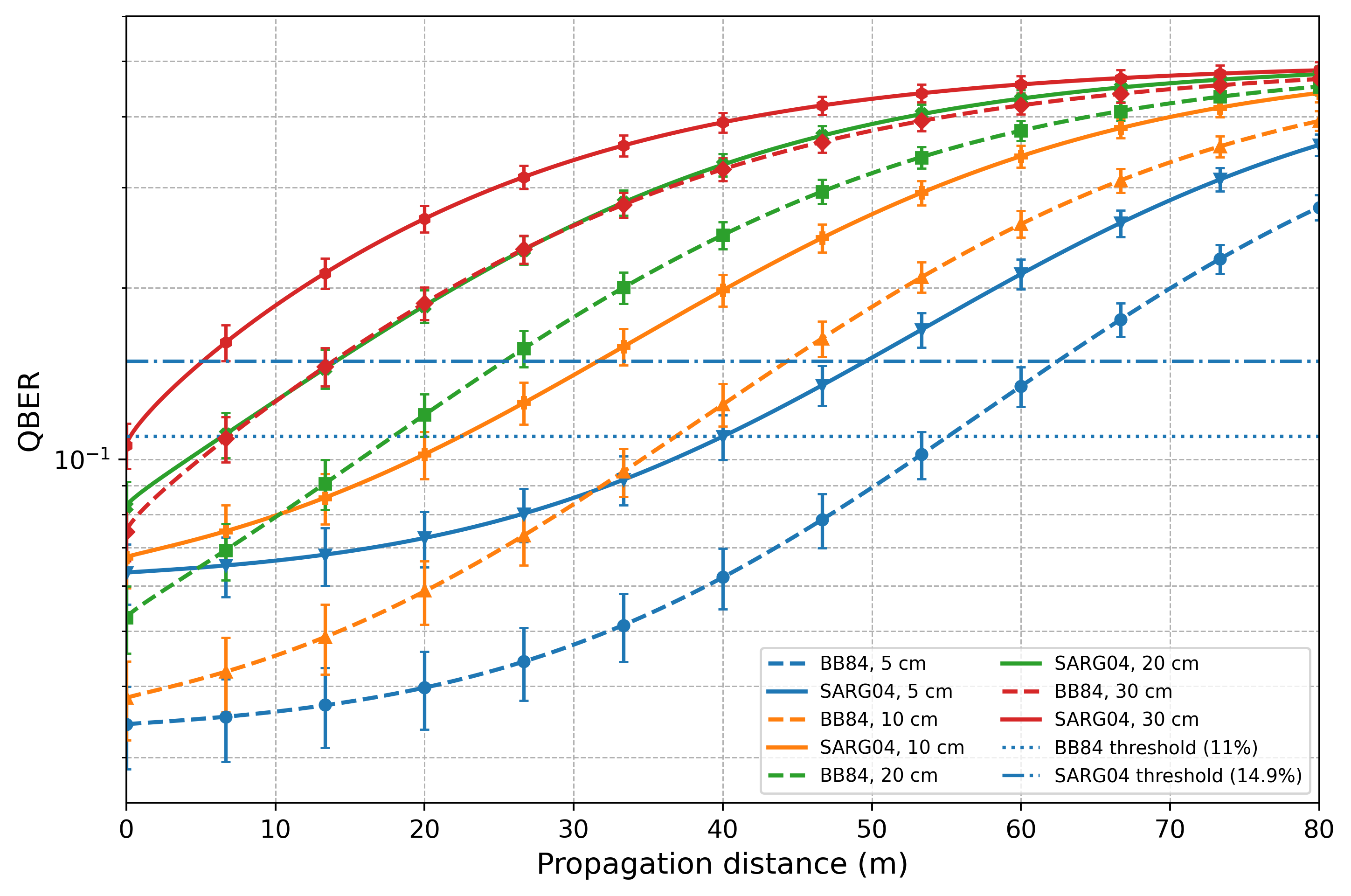}
        \\ (a)
    \end{minipage}
    \hfill
    \begin{minipage}{0.49\textwidth}
        \centering
        \includegraphics[width=\linewidth]{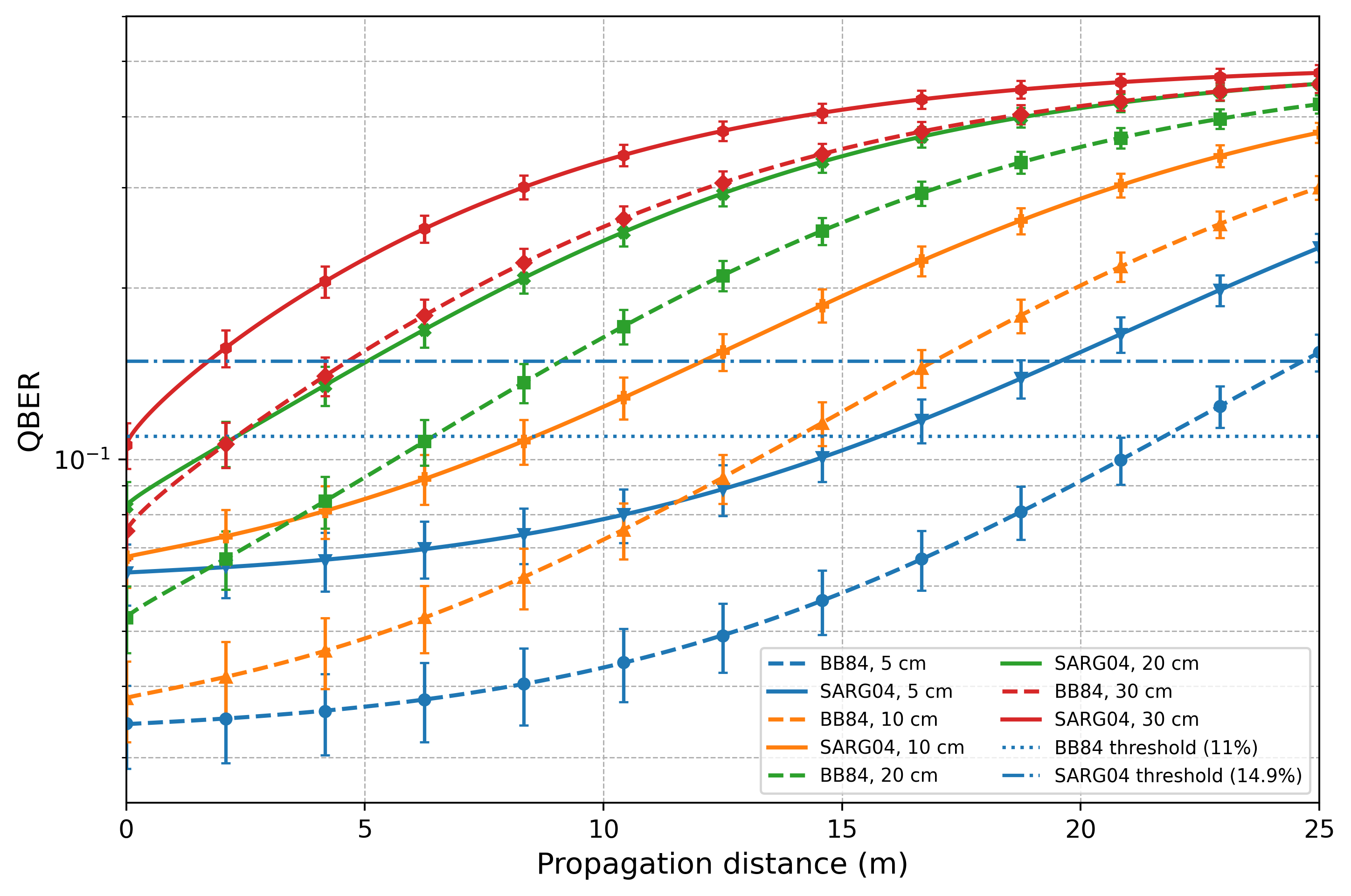}
        \\ (b)
    \end{minipage}
    \hfill
    \begin{minipage}{0.49\textwidth}
        \centering
        \includegraphics[width=\linewidth]{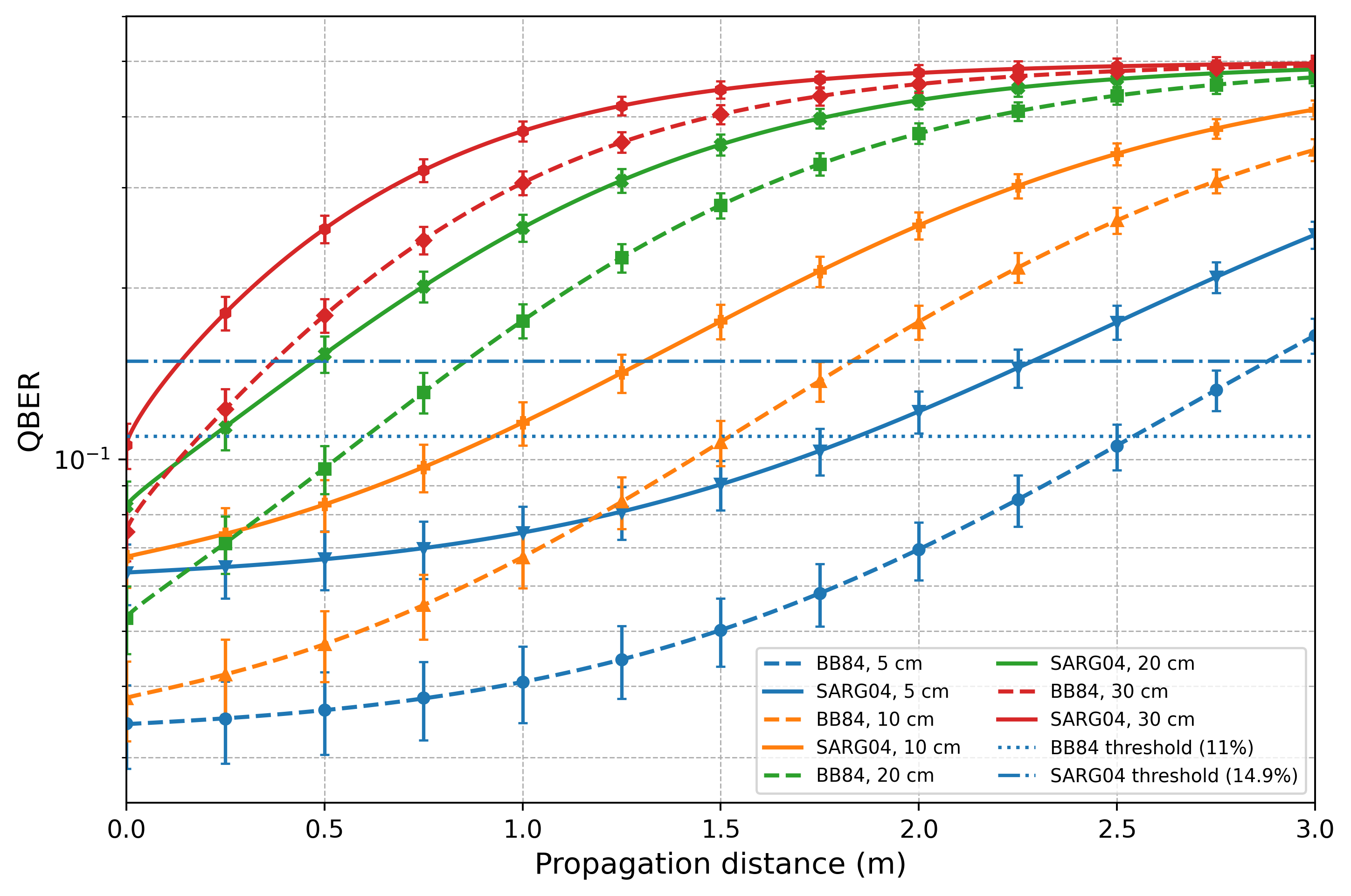}
        \\ (c)
    \end{minipage}
    
\caption{{\small
QBER$_{\mathrm{BB84}}$ and QBER$_{\mathrm{SARG04}}$ versus propagation
distance for Scenario~5, defined by the literature broadband-irradiance
benchmark
\(E_{\mathrm{lit}}=500\,\mathrm{W\,m^{-2}}\)
and mapped to the equivalent radiometric input through
Eq.~\eqref{eq:equivalent_irradiance}. The different curves correspond to
\(d_1=d_2=5\), 10, 20, and 30~cm in (a) clear, (b) coastal, and
(c) turbid water. Analytical curves follow
Eqs.~\eqref{QBER BB84} and~\eqref{QBER SARG04}; markers denote Monte Carlo
consistency estimates.
}}

\label{figure 2}
\end{figure}
{
Figures~\ref{figure 1} and~\ref{figure 2} compare the analytical QBER curves
with Monte Carlo consistency estimates for the four joint-aperture configurations
under Scenarios~1 and~5, which represent the lowest and highest literature
illumination benchmarks, respectively, calibrated via the radiometric equivalence
in Eq.~\eqref{eq:equivalent_irradiance}.

In Scenario~1 (low ambient background), expanding the joint aperture from
\(5\) to \(30~\mathrm{m}\) improves signal collection and extends the threshold
distances in clear water from \(126.34\) to \(179.15~\mathrm{m}\) for BB84, and
from \(120.06\) to \(168.72~\mathrm{m}\) for SARG04. A similar positive trend is
observed in coastal water. Conversely, in turbid water, enlarging the aperture
slightly degrades performance, reducing the BB84 and SARG04 threshold distances
from \(5.83\) to \(4.81~\mathrm{m}\) and from \(5.54\) to \(4.53~\mathrm{m}\),
respectively. Under the high illumination of Scenario~5, increasing the aperture
systematically shortens the threshold distance across all water types; in clear
water, for instance, the maximum distances drop sharply from \(55.15\) to
\(6.93~\mathrm{m}\) for BB84 and from \(49.63\) to \(5.16~\mathrm{m}\) for SARG04.

These contrasting behaviors stem from a competing balance within the joint
transmitter--receiver configuration: the aperture diameter \(d_1\) and the empirical
exponent \(T\) govern useful signal transmittance, whereas the receiver pupil area
\(S_{\rm rx} \propto d_2^2\) dictates background-noise collection. Consequently,
signal collection dominates in clear and coastal waters under low ambient light,
whereas background and scattering noise dominate under strong illumination or
severe turbidity. Across all configurations, SARG04 reaches its threshold at a
shorter distance than BB84 due to its lower conclusive-event probability. Importantly,
this outcome reflects a QBER-threshold comparison under the adopted link model rather
than a universal security hierarchy, and the reported distances represent QBER-crossing
indicators rather than operational secret-key-rate bounds.
}

\begin{figure}[!t]
    \centering
    
    \begin{minipage}{0.49\textwidth}
    
        \centering
        \includegraphics[width=\linewidth]{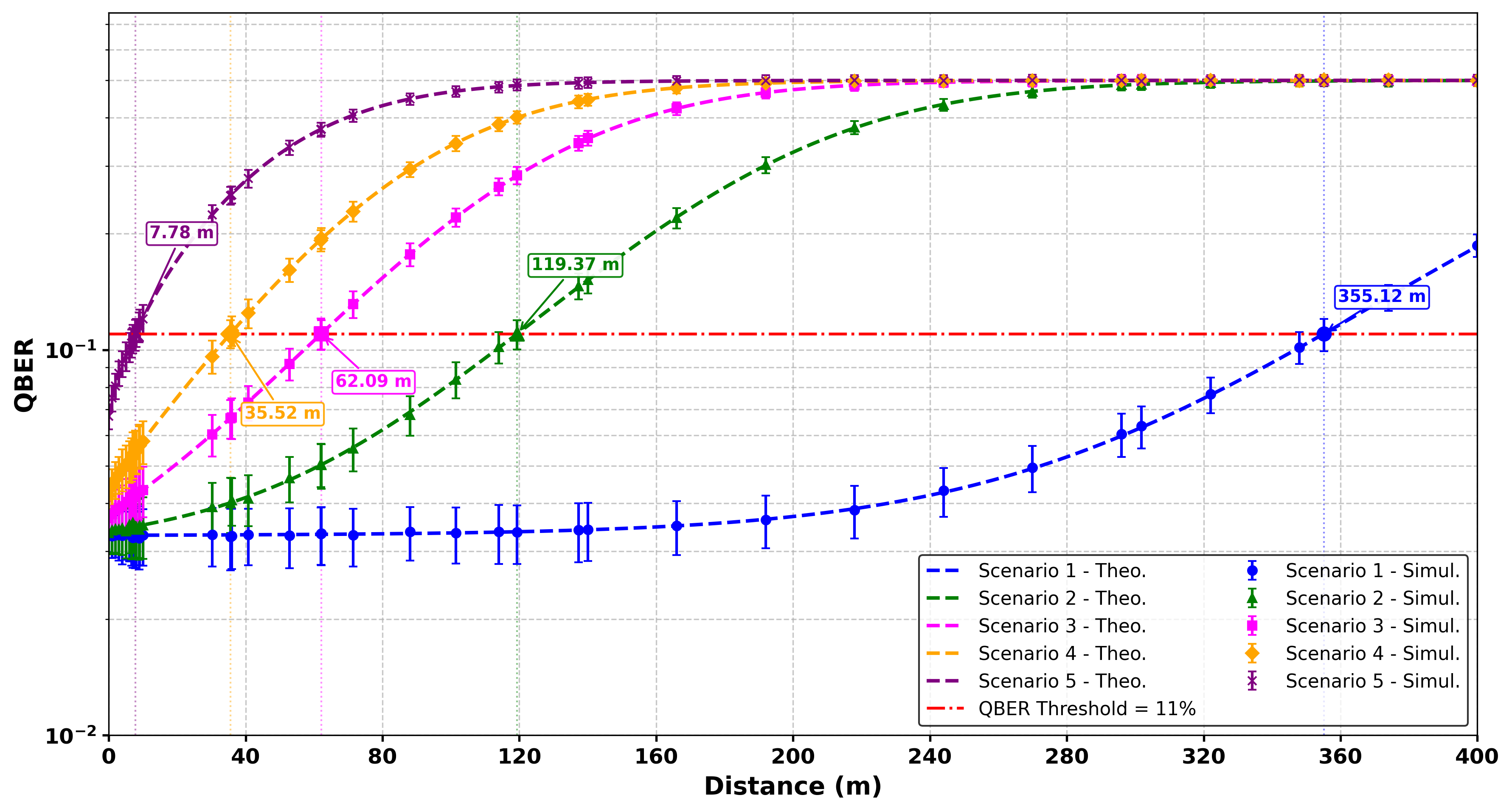}
        \\ (a)
    \end{minipage}
    \hfill
    \begin{minipage}{0.49\textwidth}
    
        \centering
        \includegraphics[width=\linewidth]{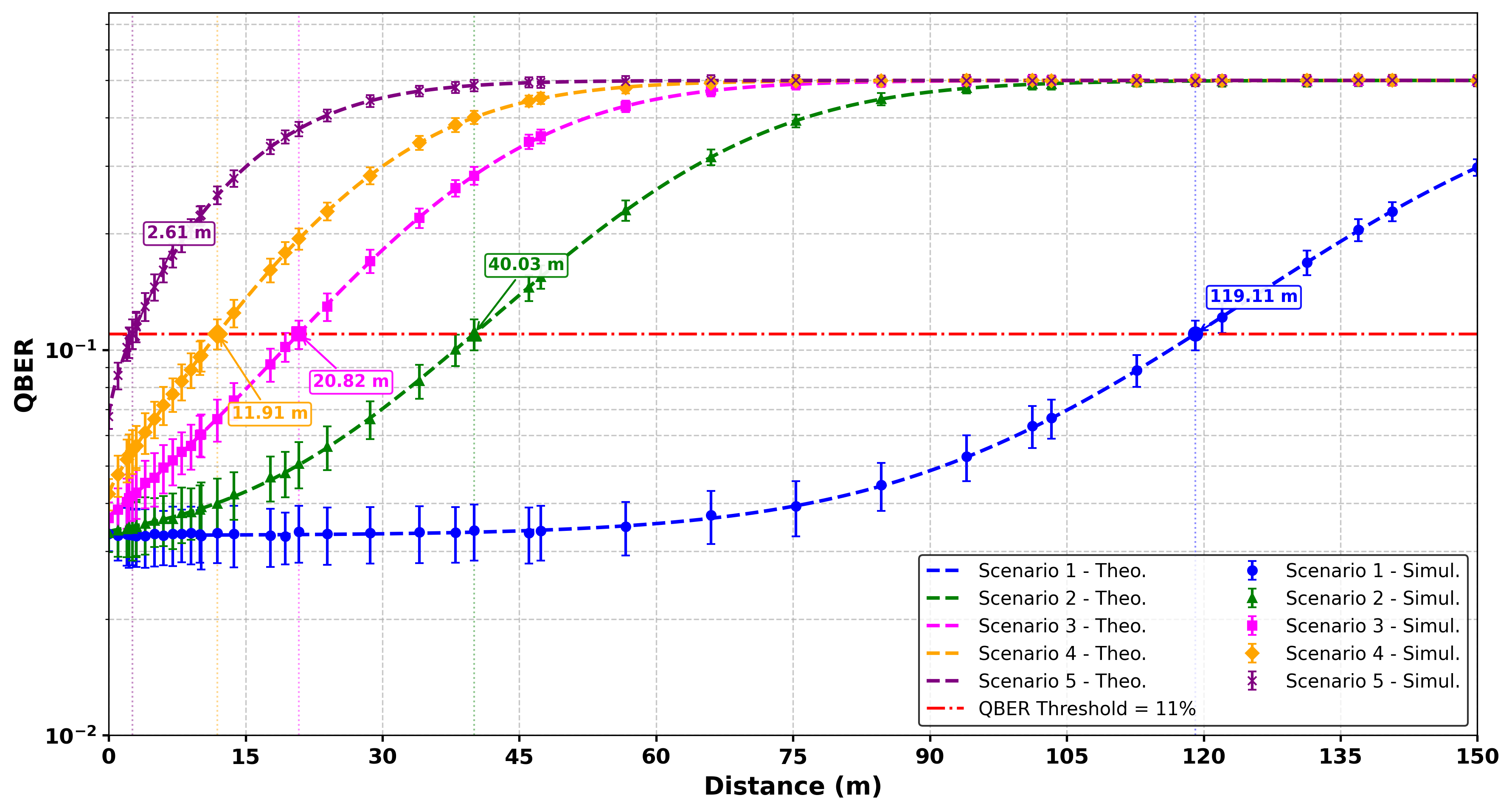}
        \\ (b)
    \end{minipage}
    \hfill
    \begin{minipage}{0.49\textwidth}
    
        \centering
        \includegraphics[width=\linewidth]{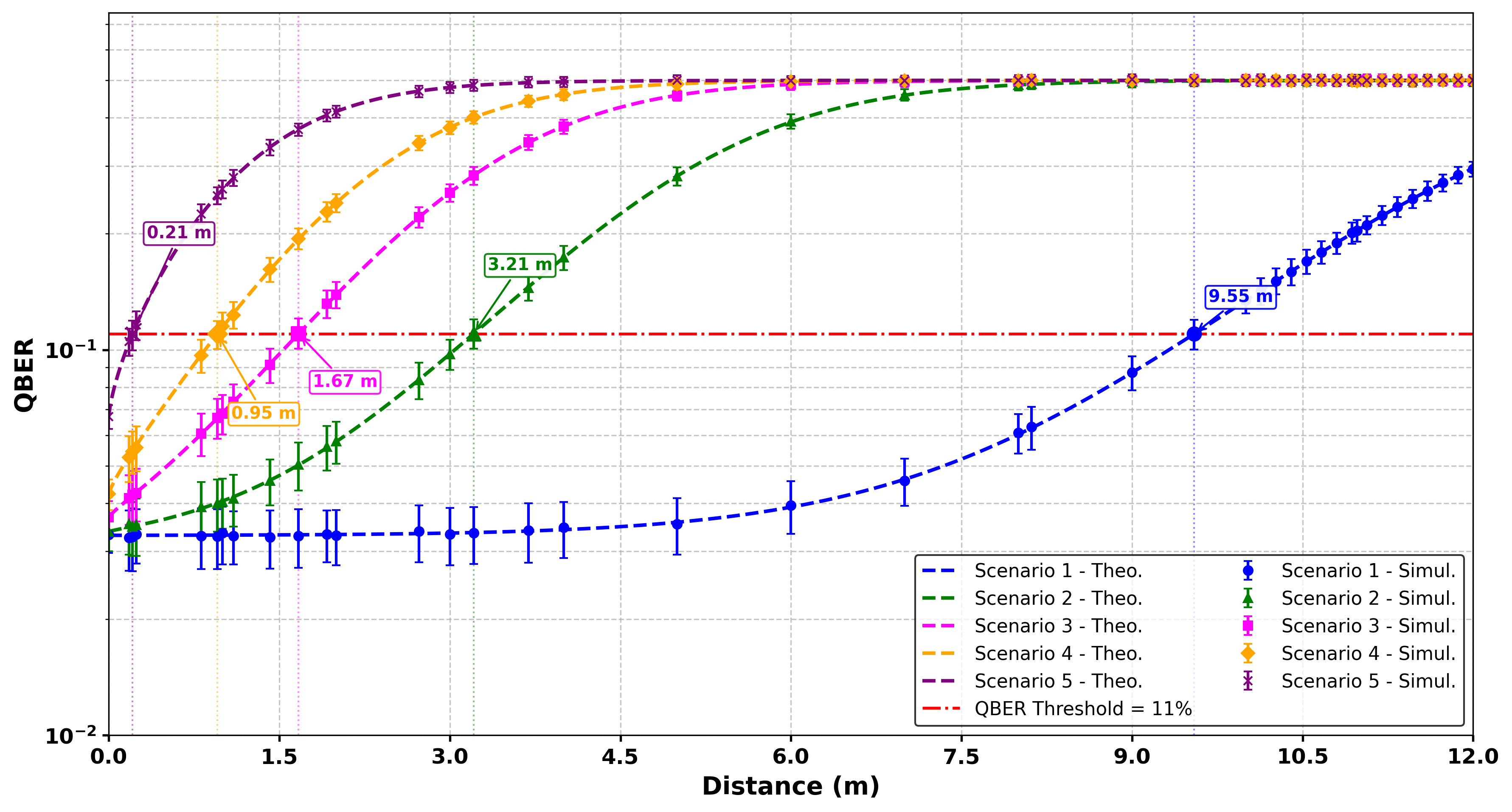}
        \\ (c)
    \end{minipage}
    
\caption{{\small
BBM92 QBER versus the total two-arm propagation length \(L\) for the five
illumination scenarios defined in
Eq.~\eqref{eq:surface_irradiance_scenarios}, with the entangled-photon
source at \(x=L/2\), in (a) clear, (b) coastal, and (c) turbid water.
Analytical curves follow Eq.~\eqref{QBER BBM92}; markers denote the
event-based Monte Carlo consistency estimates. The horizontal line indicates
the \(11\%\) QBER threshold, and the annotations identify the corresponding
threshold-crossing distances.
}}

\label{figure 4}
\end{figure}

{
Figure~\ref{figure 4} presents the analytical and Monte Carlo BBM92 QBER
as a function of the total two-arm propagation length \(L\) for a symmetric
midpoint source (\(x=L/2\)) across all five illumination scenarios. The close
agreement between the analytical curves and stochastic estimates confirms
the numerical accuracy of the implementation.

As expected, the maximum link length below the \(11\%\) QBER threshold
shrinks markedly with increasing ambient background and water turbidity.
Under the low-background conditions of Scenario~1, the threshold distances
reach approximately \(355.12~\mathrm{m}\), \(119.11~\mathrm{m}\), and
\(9.55~\mathrm{m}\) in clear, coastal, and turbid waters, respectively.
Under the severe illumination of Scenario~5, these maximum lengths drop to
\(7.78~\mathrm{m}\), \(2.61~\mathrm{m}\), and \(0.21~\mathrm{m}\).

In our decoupled modeling framework, optical attenuation primarily reduces
the true-coincidence probability, whereas the conditional polarization error
is governed independently by the local depolarizing channels. Consequently,
the sharp QBER degradation observed at long distances is driven by the drop
in signal-to-noise ratio as false coincidences become significant relative
to true coincidences.

Crucially, crossing the QBER threshold must be distinguished from achieving a
practically viable detection rate. To illustrate this distinction, consider
the clear-water Scenario~1 threshold at \(L=355.12~\mathrm{m}\). For the
idealized single-pair benchmark (\(p_{\mathrm{pair}}=1\)), the analytical model
yields a coincidence probability per \(40~\mathrm{ns}\) emission window of
\[
P_{\mathrm{coinc}}
\simeq
3.25\times10^{-9}.
\]
Assuming a source repetition rate of \(f_{\mathrm{rep}}=25~\mathrm{MHz}\),
the corresponding raw coincidence rate is
\[
R_{\mathrm{coinc}}
=
f_{\mathrm{rep}}P_{\mathrm{coinc}}
\simeq
8.11\times10^{-2}~\mathrm{s^{-1}}.
\]
Accounting for the \(1/2\) basis-matching factor, the sifted coincidence rate
reduces to
\[
R_{\mathrm{sift}}
=
\frac{1}{2}R_{\mathrm{coinc}}
\simeq
4.06\times10^{-2}~\mathrm{s^{-1}},
\]
which corresponds to approximately one sifted coincidence every \(25~\mathrm{s}\).
Thus, a link distance satisfying the QBER threshold may nevertheless yield an
extremely low usable-event throughput. Importantly, these figures reflect raw
sifted coincidences rather than secure secret-key rates, and they would decrease
even further under realistic multi-pair source statistics (\(p_{\mathrm{pair}}<1\)).
}

\begin{figure}[!t]
    \centering
    
    \begin{minipage}{0.49\textwidth}
    
        \centering
        \includegraphics[width=\linewidth]{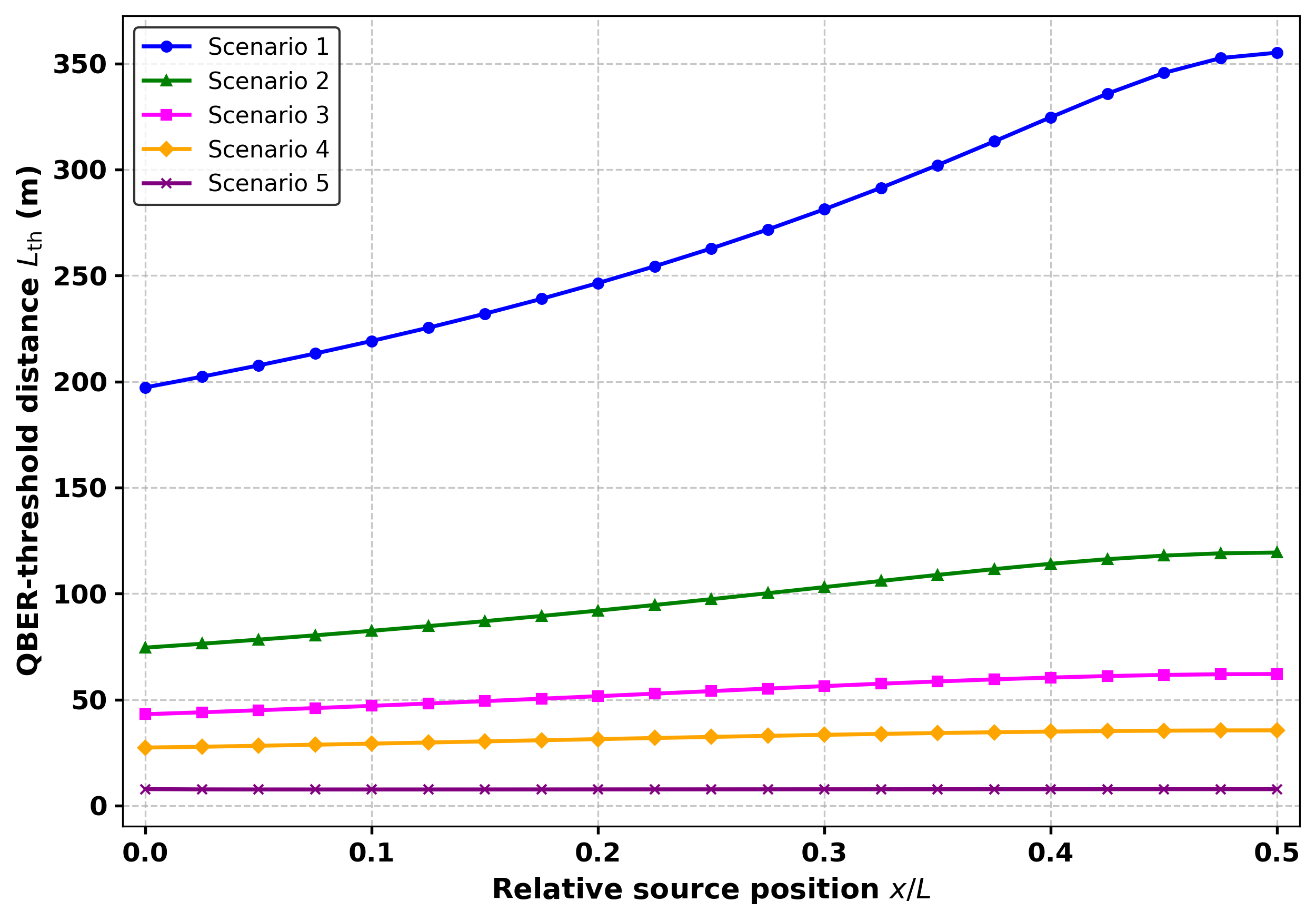}
        \\ (a)
    \end{minipage}
    \hfill
    \begin{minipage}{0.49\textwidth}
    
        \centering
        \includegraphics[width=\linewidth]{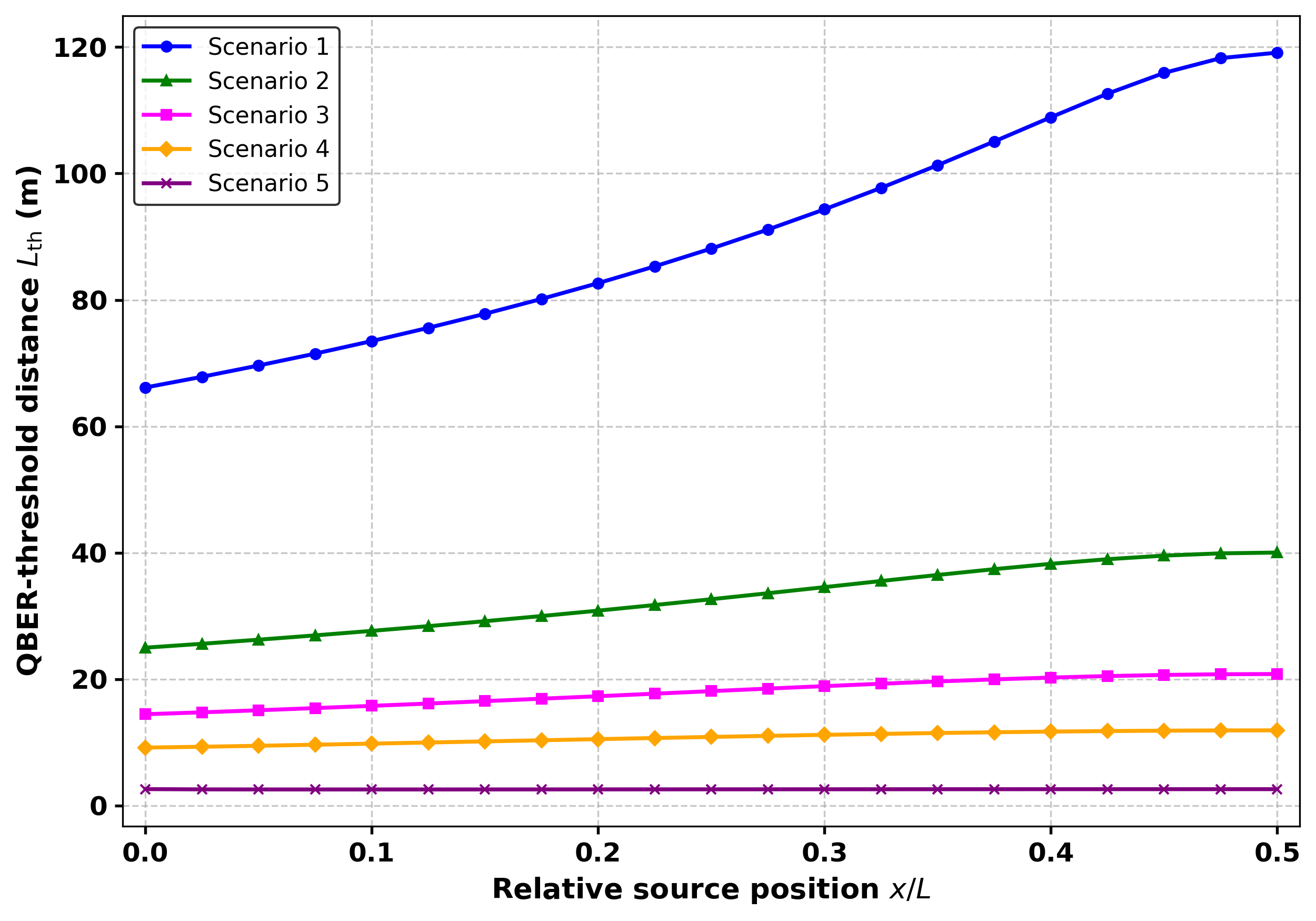}
        \\ (b)
    \end{minipage}
    \caption{{\small QBER-threshold distance \(L_{\mathrm{th}}\), defined by \(\text{QBER}_{\text{BBM92}}=11\%\), versus the relative position of the entangled source \(x/L\) in (a) clear water and (b) coastal water.}}
    \label{figure 5}
\end{figure}

{
Figure~\ref{figure 5} shows the BBM92 QBER-threshold distance
\(L_{\mathrm{th}}\) as a function of the relative source position \(x/L\).
In clear and coastal waters, the midpoint placement (\(x=L/2\)) achieves the
largest threshold distance across Scenarios~1--4. Under the strongly
background-dominated regime of Scenario~5, however, the spatial dependence
becomes shallow and slightly nonmonotonic, yielding only marginal
differences between the symmetric midpoint and asymmetric end-source
configurations. Consequently, while symmetric source placement is optimal
over most of the investigated parameter space, it does not constitute a
universal optimum. As demonstrated in Appendix~\ref{Appendix B}, this
behavior stems from a competing balance between minimizing the conditional
polarization error and preserving a high true-coincidence fraction.
}

\begin{figure}[!t]
    \centering
    
    \begin{minipage}{0.49\textwidth}
    
        \centering
        \includegraphics[width=\linewidth]{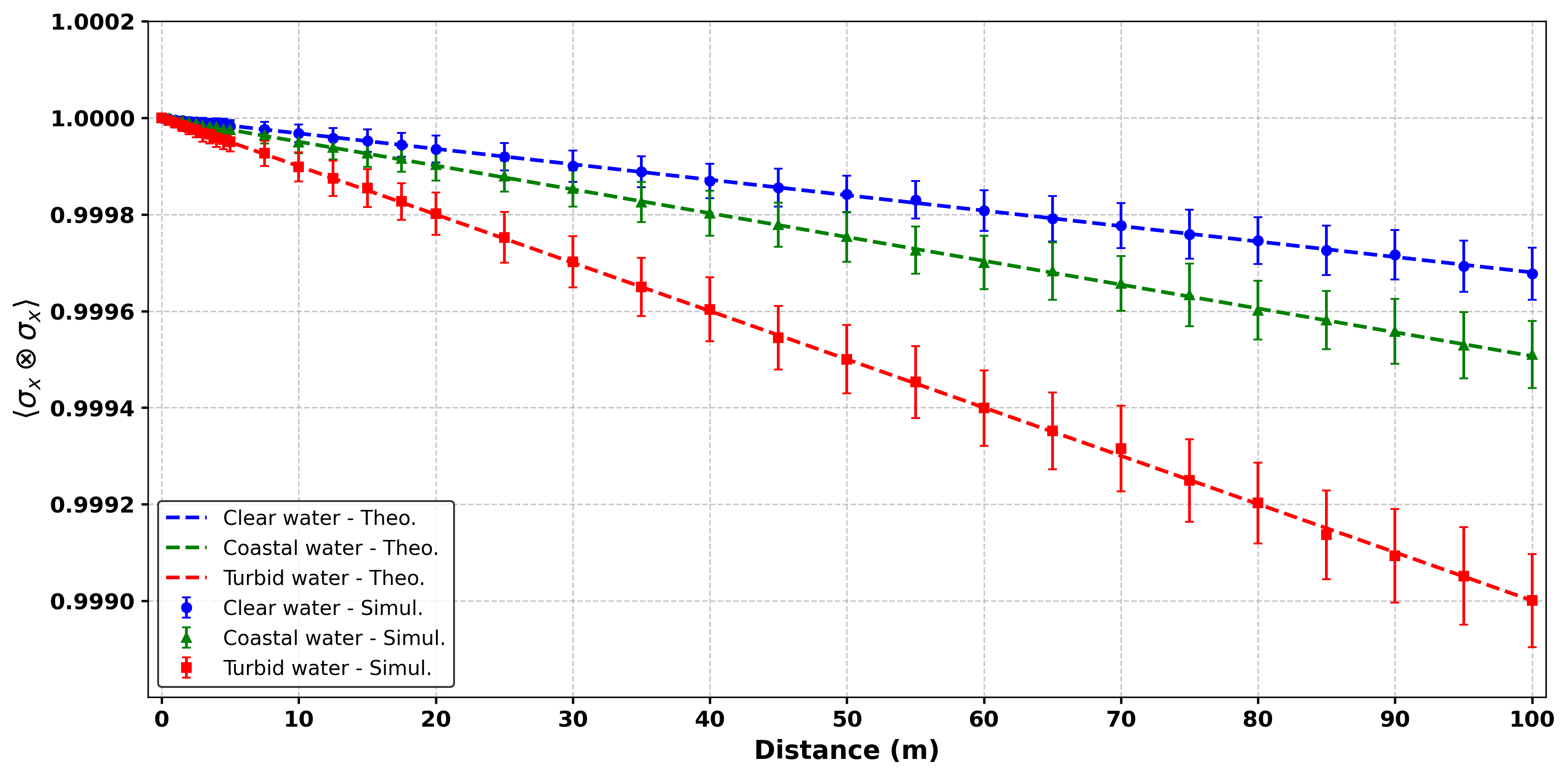}
        \\ (a)
    \end{minipage}
    \hfill
    \begin{minipage}{0.49\textwidth}
    
        \centering
        \includegraphics[width=\linewidth]{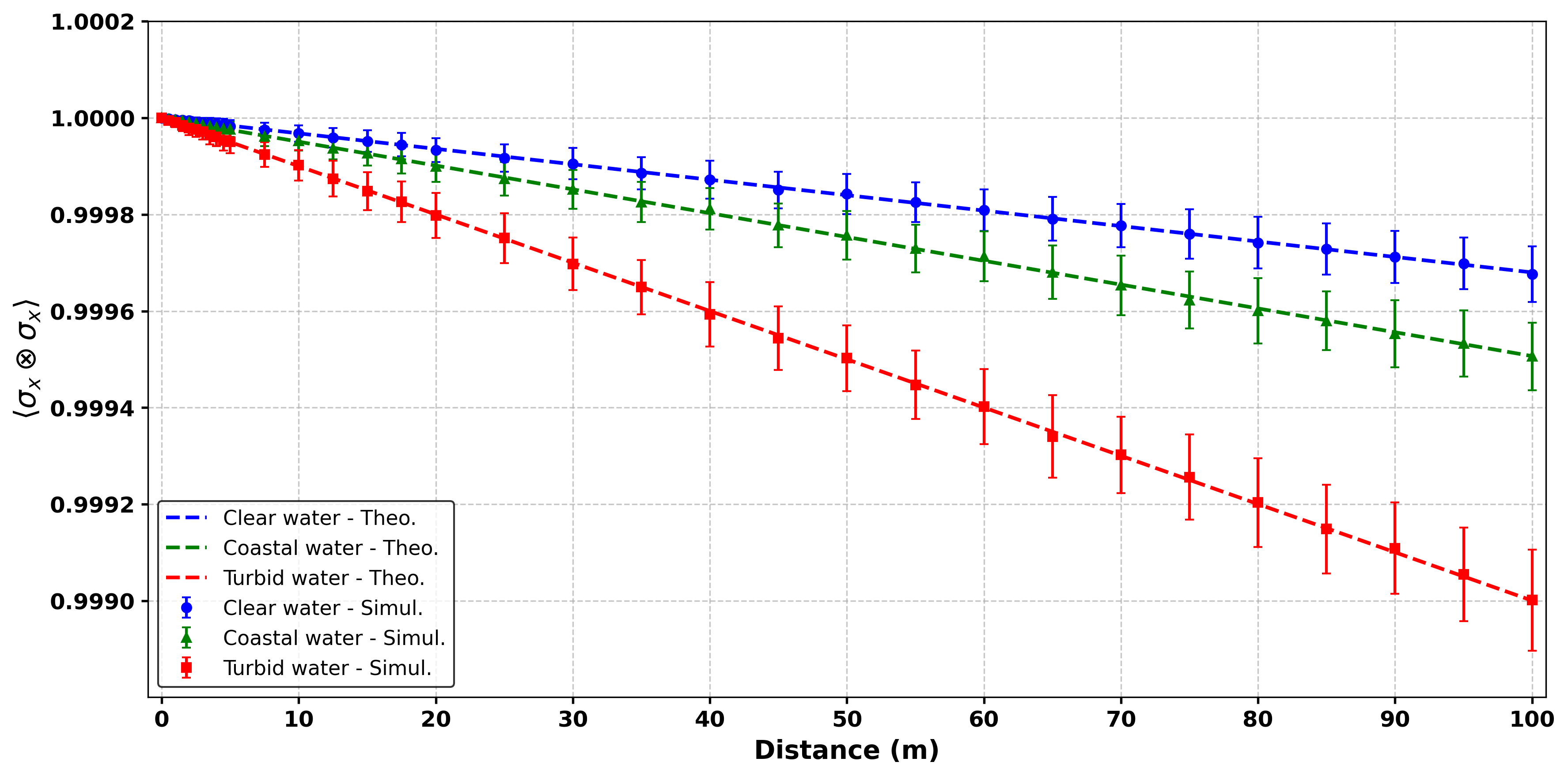}
       \\ (b)
    \end{minipage}
    
\caption{{\small
Conditional BBM92 correlation
\(C_X=\langle\sigma_x\otimes\sigma_x\rangle\) versus the total
two-arm propagation length \(L\), in clear, coastal, and turbid water,
for (a) \(x=L/2\) and (b) \(x=0.2L\). Analytical curves follow the
local-depolarization expression
\(C_X=(1-4q_A/3)(1-4q_B/3)\); markers denote the Monte Carlo
consistency estimates.
}}

\label{figure 6}
\end{figure}

{
Figure~\ref{figure 6} illustrates the conditional \(X\)-basis correlation
\(C_X = \langle\sigma_x\otimes\sigma_x\rangle\) as a function of the total
propagation length \(L\) for both symmetric (\(x=L/2\)) and asymmetric
(\(x=0.2L\)) source placements. The agreement between the analytical
curves and the Monte Carlo estimates within statistical uncertainties
provides a final numerical consistency check of the implementation.

In accordance with the decoupled noise model, the surviving polarization
correlations in both bases satisfy
\[
C_X=C_Z=
\left(1-\frac{4q_A}{3}\right)
\left(1-\frac{4q_B}{3}\right),
\]
and are therefore governed exclusively by the local depolarization
parameters rather than by Beer-Lambert optical transmittance.
Attenuation instead acts solely to reduce the probability of recording a
coincidence detection. With the effective depolarization coefficients adopted
here, the conditional correlations remain close to unity over the entire
evaluated range, showing their most pronounced reduction in turbid water.
Furthermore, symmetric midpoint placement consistently retains a correlation
equal to or slightly higher than the asymmetric configuration, in full
agreement with the convexity analysis detailed in Appendix~\ref{Appendix B}.
}

\section{Conclusion}
\label{Conclusion}
{
This study presented a unified comparative analysis of the QBER performance
of the BB84, SARG04, and BBM92 protocols across non-turbulent underwater optical
channels, evaluating diverse water turbidity types, ambient-background
illumination levels, and system configurations. For the prepare-and-measure
BB84 and SARG04 schemes, the achievable distance below the QBER threshold is
primarily dictated by channel transmittance and background-induced false clicks.
Under the assumptions evaluated here, SARG04 consistently exhibits a higher QBER
and a shorter threshold distance than BB84 due to its lower conclusive-event probability.

For the entanglement-based BBM92 protocol, optical attenuation and conditional
polarization degradation were modeled as strictly decoupled mechanisms. Photon loss
enters exclusively through the two-arm detection efficiencies to govern coincidence
rates, whereas the polarization state conditioned on successful two-photon detection
is degraded independently by local depolarizing Kraus channels. The resulting BBM92
QBER was derived from an exact algebraic partition between true and false coincidences
and from the surviving Pauli correlations in both retained measurement bases,
\(C_X\) and \(C_Z\).

Our findings demonstrate that a low conditional BBM92 QBER can persist even when
the two-photon coincidence probability has become vanishingly small. Consequently,
the maximum propagation distance below a QBER threshold does not, by itself,
define an operational communication range or a complete secret-key-security bound.
Furthermore, symmetric midpoint source placement maximizes the threshold distance
over most investigated conditions, whereas this spatial dependence flattens in
strongly background-dominated regimes.

Stochastic Monte Carlo simulations provided rigorous implementation-consistency
checks of the analytical models, without constituting independent physical validations
of the underwater channel assumptions. The scope of the present conclusions remains
confined to non-turbulent propagation, an idealized single-pair source benchmark,
and asymptotic QBER criteria. Future operational assessments of underwater QKD
will necessitate incorporating multi-pair emission statistics, coincidence throughput
constraints, finite-key composable security, and experimentally calibrated
polarization-degradation parameters.
}

\appendices
\numberwithin{equation}{section}
\section{Quantum Gain for BB84}
\label{Appendix A}
Using \eqref{gain_quantique_global} 
and the Poissonian distribution of weak coherent pulses, we obtain:
\begin{align}
&Q_{\mu, \text{BB84}} = \sum_{i=0}^{\infty} Y_{i}^{\text{(BB84)}} 
   \cdot \frac{\mu^i}{i!} e^{-\mu}, \notag \\
&= \sum_{i=1}^{\infty} \left[1 - (1 - \eta)^i + y_0 \right] 
   \cdot \frac{\mu^i}{i!} e^{-\mu} + y_0 \cdot e^{-\mu}, \notag \\
&= y_0 \cdot \left( \sum_{i=1}^{\infty} \frac{\mu^i}{i!} e^{-\mu} + e^{-\mu} \right) 
   + \sum_{i=1}^{\infty} \left[1 - (1 - \eta)^i \right] 
   \cdot \frac{\mu^i}{i!} e^{-\mu}, \notag \\
&= y_0 + \sum_{i=1}^{\infty} 
   \left[1 - (1 - \eta)^i \right] \cdot \frac{\mu^i}{i!} e^{-\mu},\notag \\
&= y_0 +  \sum_{i=1}^{\infty} \frac{\mu^i}{i!}e^{-\mu}
     - \sum_{i=1}^{\infty}(1-\eta)^i \frac{\mu^i}{i!}e^{-\mu}, \notag \\
&= y_0 + (e^{\mu}-1)e^{-\mu} - \big(e^{(1-\eta)\mu}-1\big)e^{-\mu}, \notag \\
&= y_0 + 1 - e^{-\eta\mu}.
\end{align}
\section{Source-Position Dependence of the BBM92 QBER}
\label{Appendix B}

{
This appendix examines the dependence of the exact BBM92 QBER on the
position \(x\) of the entangled-photon source. For identical receivers
in a homogeneous water type, the two arm efficiencies are
\begin{equation}
\eta_A(x)
=
\eta_0 A(x,\lambda),
\qquad
\eta_B(x)
=
\eta_0 A(L-x,\lambda),
\label{eq:appendix_position_efficiencies}
\end{equation}
where \(\eta_0=\eta_{\mathrm{Alice}}=\eta_{\mathrm{Bob}}\). The
background-click probabilities are also assumed identical,
\(y_A=y_B=y\).

Using Eq.~\eqref{eq:BBM92_receiver_click}, the click probabilities
conditioned on pair generation are
\begin{equation}
c_A(x)
=
\eta_A(x)+[1-\eta_A(x)]y,
\qquad
c_B(x)
=
\eta_B(x)+[1-\eta_B(x)]y.
\label{eq:appendix_position_clicks}
\end{equation}
The exact true-coincidence and total-coincidence probabilities are therefore
\begin{align}
P_{\mathrm{true}}(x)
&=
p_{\mathrm{pair}}\eta_A(x)\eta_B(x),
\notag\\
P_{\mathrm{coinc}}(x)
&=
p_{\mathrm{pair}}c_A(x)c_B(x)
+
(1-p_{\mathrm{pair}})y^2.
\label{eq:appendix_position_coincidences}
\end{align}

Since
\(P_{\mathrm{false}}=P_{\mathrm{coinc}}-P_{\mathrm{true}}\),
Eq.~\eqref{QBER BBM92} can be rewritten as
\begin{equation}
\mathrm{QBER}_{\mathrm{BBM92}}(x)
=
\frac{1}{2}
-
\left[
\frac{1}{2}-e_{\mathrm{sig}}(x)
\right]
\frac{P_{\mathrm{true}}(x)}
     {P_{\mathrm{coinc}}(x)}.
\label{eq:appendix_position_qber}
\end{equation}
This form separates the two mechanisms governing the QBER. The factor
\(1/2-e_{\mathrm{sig}}\) describes the quality of the surviving
polarization correlations, whereas
\(P_{\mathrm{true}}/P_{\mathrm{coinc}}\) is the fraction of recorded
coincidences that originate from the two photons of the same pair.

For the local depolarizing model,
\[
q(\ell)=1-\exp(-\gamma_{\mathrm{dep}}\ell),
\]
the correlation-contraction factor is
\begin{align}
r(x)
&=
\left[
1-\frac{4q(x)}{3}
\right]
\left[
1-\frac{4q(L-x)}{3}
\right]
\notag\\
&=
\frac{
\left[4e^{-\gamma_{\mathrm{dep}}x}-1\right]
\left[4e^{-\gamma_{\mathrm{dep}}(L-x)}-1\right]
}{9}
\notag\\
&=
\frac{
16e^{-\gamma_{\mathrm{dep}}L}
-4\left[
e^{-\gamma_{\mathrm{dep}}x}
+e^{-\gamma_{\mathrm{dep}}(L-x)}
\right]
+1
}{9}.
\label{eq:appendix_position_r}
\end{align}
Because
\[
e^{-\gamma_{\mathrm{dep}}x}
+
e^{-\gamma_{\mathrm{dep}}(L-x)}
\geq
2e^{-\gamma_{\mathrm{dep}}L/2},
\]
with equality at \(x=L/2\), it follows that
\begin{equation}
r(x)\leq r(L/2).
\end{equation}
The midpoint therefore minimizes the conditional signal-error probability
\(e_{\mathrm{sig}}\).

The complete QBER also depends on the true-coincidence fraction in
Eq.~\eqref{eq:appendix_position_qber}. Under the transformation
\(x\leftrightarrow L-x\), the two arm efficiencies and click probabilities
are exchanged, while their products remain unchanged. Consequently,
\begin{equation}
\mathrm{QBER}_{\mathrm{BBM92}}(x)
=
\mathrm{QBER}_{\mathrm{BBM92}}(L-x),
\label{eq:appendix_position_symmetry}
\end{equation}
and, whenever the derivative exists,
\begin{equation}
\left.
\frac{\partial\,\mathrm{QBER}_{\mathrm{BBM92}}}
     {\partial x}
\right|_{x=L/2}
=
0.
\end{equation}

Equation~\eqref{eq:appendix_position_symmetry} establishes that the
midpoint is a stationary configuration, but it does not by itself prove a
universal global minimum. The exact position dependence also involves the
modified Beer--Lambert transmittances in
\(P_{\mathrm{true}}/P_{\mathrm{coinc}}\). When the recorded coincidences
are predominantly true coincidences, balancing the two arms and maximizing
the conditional correlation favors the midpoint. When the background
contribution dominates, the position dependence becomes shallow and a
slightly asymmetric configuration may give a marginally different
QBER-threshold distance.

The global source-position dependence reported in Fig.~\ref{figure 5} is
therefore obtained by direct numerical evaluation of the complete
Eq.~\eqref{eq:appendix_position_qber}, rather than by minimizing the false
coincidence probability alone.
}

\section{
\texorpdfstring{
{Derivation of the Conditional BBM92 State Under Local Depolarizing Channels}
}{
Derivation of the Conditional BBM92 State Under Local Depolarizing Channels
}
}
\label{appendix:BBM92_depolarization}

{
This appendix derives the conditional polarization state used in
Eq.~\eqref{equa rho depo}. Photon loss does not appear in this derivation,
because it is already included through the arm detection efficiencies
\(\eta_A\) and \(\eta_B\). The derivation below is therefore conditioned on
the successful detection of both photons.

For each propagation arm \(X\in\{A,B\}\), the four local Kraus operators
can be written as
\[
L_i(q_X)
=
\sqrt{w_i^{(X)}}\,\sigma_i,
\qquad
i\in\{0,1,2,3\},
\]
where
\[
\sigma_0=I,
\qquad
\sigma_1=\sigma_x,
\qquad
\sigma_2=\sigma_y,
\qquad
\sigma_3=\sigma_z,
\]
and
\[
w_0^{(X)}=1-q_X,
\qquad
w_1^{(X)}
=
w_2^{(X)}
=
w_3^{(X)}
=
\frac{q_X}{3}.
\]
The local and global completeness relations are
\begin{align}
\sum_{i=0}^{3}
L_i^{\dagger}(q_X)L_i(q_X)
&=
\left[
1-q_X
+
3\left(\frac{q_X}{3}\right)
\right]I
=
I,
\notag\\
\sum_{i,j=0}^{3}
L_{ij}^{\dagger}L_{ij}
&=
\left(
\sum_{i=0}^{3}
L_i^{\dagger}(q_A)L_i(q_A)
\right)
\otimes
\left(
\sum_{j=0}^{3}
L_j^{\dagger}(q_B)L_j(q_B)
\right)
\notag\\
&=
I_A\otimes I_B,
\label{eq:appendix_global_completeness}
\end{align}
where
\[
L_{ij}
=
L_i(q_A)\otimes L_j(q_B).
\]
The global map is therefore completely positive and trace preserving.

Using the weights defined above, the explicit sum over the 16 global
Kraus terms is
\begin{align}
\rho_{AB}^{\mathrm{out-dep}}
&=
\sum_{i,j=0}^{3}
w_i^{(A)}w_j^{(B)}
\left(\sigma_i\otimes\sigma_j\right)
\rho_{AB}^{\mathrm{cond}}
\left(\sigma_i\otimes\sigma_j\right).
\label{eq:appendix_sixteen_kraus_sum}
\end{align}

The conditional Bell state can be expressed in the two-qubit Pauli basis
as
\begin{equation}
\rho_{AB}^{\mathrm{cond}}
=
\ket{\Phi^+}\bra{\Phi^+}
=
\frac{1}{4}
\left(
I\otimes I
+
\sigma_x\otimes\sigma_x
-
\sigma_y\otimes\sigma_y
+
\sigma_z\otimes\sigma_z
\right).
\label{eq:appendix_bell_pauli_decomposition}
\end{equation}

For \(k\in\{x,y,z\}\), the action of one local depolarizing channel on a
Pauli operator is
\begin{align}
\mathcal{E}_{X}(\sigma_k)
&=
(1-q_X)\sigma_k
+
\frac{q_X}{3}
\sum_{i=x,y,z}
\sigma_i\sigma_k\sigma_i
\notag\\
&=
\left(
1-q_X
+
\frac{q_X}{3}
-
\frac{2q_X}{3}
\right)\sigma_k
\notag\\
&=
\left(1-\frac{4q_X}{3}\right)\sigma_k
=
\kappa_X\sigma_k,
\label{eq:appendix_local_pauli_contraction}
\end{align}
where
\[
\kappa_X
=
1-\frac{4q_X}{3}.
\]
The identity operator remains unchanged,
\(\mathcal{E}_X(I)=I\).

Applying the two local channels to
Eq.~\eqref{eq:appendix_bell_pauli_decomposition} therefore gives
\begin{align}
\rho_{AB}^{\mathrm{out-dep}}
&=
\frac{1}{4}
\left[
I\otimes I
+
r\,\sigma_x\otimes\sigma_x
-
r\,\sigma_y\otimes\sigma_y
+
r\,\sigma_z\otimes\sigma_z
\right]
\notag\\
&=
\frac{1}{4}
\begin{pmatrix}
1+r & 0 & 0 & 2r\\
0 & 1-r & 0 & 0\\
0 & 0 & 1-r & 0\\
2r & 0 & 0 & 1+r
\end{pmatrix},
\label{eq:appendix_depolarized_state}
\end{align}
with
\begin{equation}
r
=
\kappa_A\kappa_B
=
\left(1-\frac{4q_A}{3}\right)
\left(1-\frac{4q_B}{3}\right).
\label{eq:appendix_total_contraction}
\end{equation}

The corresponding Pauli correlations are consequently
\begin{equation}
C_X=r,
\qquad
C_Y=-r,
\qquad
C_Z=r.
\label{eq:appendix_correlations}
\end{equation}
The eigenvalues of
\(\rho_{AB}^{\mathrm{out-dep}}\) are
\[
\frac{1+3r}{4},
\qquad
\frac{1-r}{4},
\qquad
\frac{1-r}{4},
\qquad
\frac{1-r}{4},
\]
which confirms the positivity of the density matrix for the values of
\(r\) generated by the local channels. Its trace is equal to one.

This derivation shows explicitly that the local depolarizing channels
contract the conditional \(X\)- and \(Z\)-basis correlations by the same
factor \(r\). Underwater attenuation instead acts through the probability
of recording a two-photon coincidence and is not applied again to the
conditional polarization state.
}

\section{
\texorpdfstring{
{Effective Depolarization Calibration at 810 nm}
}{
Effective Depolarization Calibration at 810 nm
}
}
The approach proposed in \cite{ji2017towards} consists of comparing quantum correlation measured before and after transmission through a 3 m underwater channel. In an ideal Bell state, the observable 
\(\langle \sigma_x \otimes \sigma_x \rangle\) takes its maximum value in the absence of noise, meaning that the information
is perfectly conserved. Conversely, when one of the photons passes through a depolarizing channel, this correlation decreases. We then define the ratio
\begin{equation}
R = \frac{V_{\text{out}}}{V_{\text{in}}},
\end{equation}
where \(V_{\text{in}} = \mathrm{Tr}\!\left(\rho_{AB}^{\text{in}} 
(\sigma_x \otimes \sigma_x)\right)\) and \(V_{\text{out}}\) represent the correlations before and after depolarization, respectively. To explicitly compute \(V_{\text{out}}\), we apply the action of the depolarizing channel to the correlation
{\small
\begin{align}
V_{\text{out}} &= (1 - q) V_{\text{in}} 
+ \frac{q}{3} \!\!\!\sum_{i=x,y,z}\!\!\!\!
\text{Tr}\!\left[(I_A \otimes \sigma_i) \rho_{AB}^{\text{in}} (I_A \otimes \sigma_i) (\sigma_x \otimes \sigma_x)\right], \notag\\
&= (1 - q) V_{\text{in}} 
+ \frac{q}{3} \sum_{i=x,y,z}\!\!
\text{Tr}\!\left[\rho_{AB}^{\text{in}} \big(\sigma_x \otimes (\sigma_i \sigma_x \sigma_i)\big)\right],
\end{align}}
where \begin{equation*}
    \sigma_i \sigma_x \sigma_i = 
\begin{cases}
\sigma_x & \text{if } i = x, \\
-\sigma_x & \text{if } i \neq x.
\end{cases}
\end{equation*}
We deduce
\begin{equation}
q = \frac{3}{4} (1 - R).
\end{equation}
{
Using the experimental visibilities
\(V_{\mathrm{in}}=0.9528\) and \(V_{\mathrm{out}}=0.9499\), together with
the propagation length \(L=3~\mathrm{m}\), the adopted isotropic
Pauli-channel model gives
\[
R
=
\frac{V_{\mathrm{out}}}{V_{\mathrm{in}}}
\simeq 0.996956,
\qquad
q
=
\frac{3}{4}(1-R)
\simeq 2.28\times10^{-3}.
\]
Using the phenomenological relation
\[
q(L)=1-\exp\!\left(-\gamma_{\mathrm{dep}}L\right),
\]
the corresponding effective coefficient is
\[
\gamma_{\mathrm{dep}}(810~\mathrm{nm})
=
-\frac{1}{L}\ln(1-q)
\simeq
7.62\times10^{-4}~\mathrm{m^{-1}}.
\]
This value is an effective coefficient inferred within the adopted channel
model and for the specific experimental configuration of
\cite{ji2017towards}; it should not be interpreted as a universal material
constant.
}
{
Since \(q\ll1\) over the calibration distance, the phenomenological law
reduces to
\[
q(\ell)\simeq\gamma_{\mathrm{dep}}\ell.
\]
The extracted coefficient therefore characterizes the measured
first-order polarization-degradation rate within this specific experimental
and channel-model setting.
}

\section*{Acknowledgments}
The authors would like to express their gratitude to the Brittany Region for its support and funding of this research.

\label{sec:refs}
\bibliographystyle{IEEEtran}
\bibliography{biblio2}

@InProceedings{brassard1984quantum,
  author    = {C. H. Bennett and G. Brassard},
  title     = {Quantum cryptography: Public key distribution and coin tossing},
  booktitle = {International conference on computers, systems and signal processing},
  year      = {1984},
  pages     = {175--179},
  note       = {http://doi.org/10.1016/j.tcs.2014.05.025},
}

@Article{wootters1982single,
  author    = {W. K. Wootters and W. H. Zurek},
  title     = {A single quantum cannot be cloned},
  journal   = {Nature},
  year      = {1982},
  volume    = {299},
  number    = {5886},
  pages     = {802--803},
  publisher = {Nature Publishing Group UK London},
  note       = {http://doi.org/10.1038/299802a0},
}

@article{busch2007heisenberg,
  title={Heisenberg's uncertainty principle},
  author={P. Busch and T. Heinonen and P. Lahti},
  journal={Physics reports},
  volume={452},
  number={6},
  pages={155--176},
  year={2007},
  publisher={Elsevier},
  note       = {http://doi.org/10.1016/j.physrep.2007.05.006},
}

@article{scarani2004quantum,
  title={Quantum Cryptography Protocols Robust against Photon Number Splitting Attacks for Weak Laser Pulse Implementations},
  author={V. Scarani and H. Bechmann-Pasquinucci and N. J. Cerf and M. Du\v{s}ek and N. L{\"u}tkenhaus and M. Peev},
  journal={Physical review letters},
  volume={92},
  number={5},
  pages={057901},
  year={2004},
  publisher={APS},
  note= {http://doi.org/10.1103/PhysRevLett.92.057901},
}

@article{bennett1992quantum,
  title={Quantum cryptography without Bell's theorem},
  author={C. H. Bennett and G. Brassard and N. D. Mermin},
  journal={Physical review letters},
  volume={68},
  number={5},
  pages={557},
  year={1992},
  publisher={APS},
  note= {http://doi.org/10.1103/PhysRevLett.68.557},
}

@article{paglierani2023primer,
  title={A primer on underwater quantum key distribution},
  author={P. Paglierani and A. H. Fahim Raouf and K. Pelekanakis and R. Petroccia and J. Alves and M. Uysal},
  journal={Quantum Engineering},
  volume={2023},
  number={1},
  pages={7185329},
  year={2023},
  publisher={Wiley Online Library},
  note= {http://doi.org/10.1155/2023/7185329},
}

@article{waks2002security,
  title={Security of quantum key distribution with entangled photons against individual attacks},
  author={E. Waks and H. Krovi and Y. Yamamoto},
  journal={Physical Review A},
  volume={65},
  number={5},
  pages={052310},
  year={2002},
  publisher={APS},
  note= {http://doi.org/10.1103/PhysRevA.65.052310},
}

@article{shor2000simple,
  title={Simple proof of security of the {BB84} quantum key distribution protocol},
  author={P. W. Shor and J. Preskill},
  journal={Physical review letters},
  volume={85},
  number={20},
  pages={441},
  year={2000},
  publisher={APS},
  note       = {http://doi.org/10.1103/PhysRevLett.85.441},
}

@article{branciard2005security,
  title={Security of two quantum cryptography protocols using the same four qubit states},
  author={C. Branciard and N. Gisin and B. Kraus and V. Scarani},
  journal={Physical Review A},
  volume={72},
  number={3},
  pages={032301},
  year={2005},
  publisher={APS},
  note       = {http://doi.org/10.1103/PhysRevA.72.032301},
}

@article{ma2007quantum,
  title={Quantum key distribution with entangled photon sources},
  author={X. Ma and C. H. F. Fung and H. K. Lo},
  journal={Physical Review A},
  volume={76},
  number={1},
  pages={012301},
  year={2007},
  publisher={APS},
  note= {http://doi.org/10.1103/PhysRevA.76.012307},
}

@article{muskan2023analysing,
  title={Performance analysis of satellite-based {QKD} protocols using the circular beam model},
  author={Muskan and R. Meena and S. Banerjee},
  journal={arXiv},
  year={2023},
  note = {https://arxiv.org/abs/2308.01036}
}

@book{mobley1994light,
  title={Light and water: radiative transfer in natural waters},
  author={C. D. Mobley},
  year={1994},
  publisher={Academic press},
  note      = {ISBN: 978-0-12-502890-7},
}

@InProceedings{rizkperformances,
  author    = {N. Rizk and A. Dr{\'e}meau and A. Coatanhay},
  title     = {Performances des protocoles {QKD}, {BB84} et {SARG04} pour les communications quantiques sous-marines},
  booktitle = {Proc. of GRETSI, Strasbourg, France},
  year      = {2025},
}

@article{baykal2022underwater,
  title={Underwater turbulence, its effects on optical wireless communication and imaging: A review},
 author={Y. Baykal and Y. Ata and M.C. G{\"o}k{\c{c}}e},
  journal={Optics \& Laser Technology},
  volume={156},
  pages={108624},
  year={2022},
  publisher={Elsevier},
  note= {http://doi.org/10.1016/j.optlastec.2022.108624},
}

@article{bennett1989experimental,
  title={Experimental quantum cryptography},
  author={C. H. Bennett and F. Bessette and G. Brassard and L. Salvail and J. Smolin},
  journal={Journal of cryptology},
  volume={5},
  number={1},
  pages={3--28},
  year={1992},
  publisher={Springer},
  note       = {http://doi.org/10.1007/BF00191318},
}

@article{buttler1998practical,
  title={Practical free-space quantum key distribution over 1 km},
  author={W.T. Buttler and R.J. Hughes and P.G. Kwiat and S.K. Lamoreaux and G.G. Luther and G.L. Morgan and J.E. Nordholt and C.G. Peterson and C.M. Simmons},
  journal={Physical review letters},
  volume={81},
  number={15},
  pages={3283},
  year={1998},
  publisher={APS},
  note       = {http://doi.org/10.1103/PhysRevLett.81.3283},
}

@article{takesue200610,
  title={10-{GHz} clock differential phase shift quantum key distribution experiment},
  author={H. Takesue and E. Diamanti and C. Langrock and M. M. Fejer and Y. Yamamoto},
  journal={Optics Express},
  volume={14},
  number={20},
  pages={9522--9530},
  year={2006},
  publisher={Optica Publishing Group},
  note       = {http://doi.org/10.1364/oe.14.009522},
}

@article{takesue2007quantum,
  title={Quantum key distribution over a 40-{dB} channel loss using superconducting single-photon detectors},
  author={H. Takesue and S.W. Nam and Q. Zhang and R.H. Hadfield and T. Honjo and K. Tamaki and Y. Yamamoto},
  journal={Nature photonics},
  volume={1},
  number={6},
  pages={343--348},
  year={2007},
  publisher={Nature Publishing Group},
  note       = {http://doi.org/10.1038/nphoton.2007.75},
}

@article{wang20122,
  title={2 {GHz} clock quantum key distribution over 260 km of standard telecom fiber},
  author={S. Wang and W. Chen and J. F. Guo and Z. Q. Yin and H. W. Li and Z. Zhou and G. C. Guo and Z. F. Han},
  journal={Optics Letters},
  volume={37},
  number={6},
  pages={1008--1010},
  year={2012},
  publisher={Optica Publishing Group},
  note       = {http://doi.org/10.1364/OL.37.001008},
}

@article{sharma2015controlled,
  title={Controlled bidirectional remote state preparation in noisy environment: a generalized view},
  author={V. Sharma and C. Shukla and S. Banerjee and A. Pathak},
  journal={Quantum Information Processing},
  volume={14},
  number={9},
  pages={3441--3464},
  year={2015},
  publisher={Springer},
  note       = {http://doi.org/10.1007/s11128-015-1038-5},
}

@article{sidhu2021advances,
  title={Advances in space quantum communications},
  author={J.S. Sidhu and S.K. Joshi and M. G{\"u}ndo{\u{g}}an and T. Brougham and D. Lowndes and L. Mazzarella and M. Krutzik and S. Mohapatra and D. Dequal and G. Vallone and al.},
  journal={IET Quantum Communication},
  volume={2},
  number={4},
  pages={182--217},
  year={2021},
  publisher={Wiley Online Library},
  note       = {http://doi.org/10.1049/qtc2.12015},
}

@article{ecker2021strategies,
  title={Strategies for achieving high key rates in satellite-based {QKD}},
  author={S. Ecker and B. Liu and J. Handsteiner and M. Fink and D. Rauch and F. Steinlechner and T. Scheidl and A. Zeilinger and R. Ursin},
  journal={npj Quantum Information},
  volume={7},
  number={1},
  pages={5},
  year={2021},
  publisher={Nature Publishing Group UK London},
  note       = {http://doi.org/10.1038/s41534-020-00335-5},
}

@article{acosta2024analysis,
  title={Analysis of satellite-to-ground quantum key distribution with adaptive optics},
  author={V.M. Acosta and D. Dequal and M. Schiavon and A. Montmerle-Bonnefois and C.B. Lim and J.M. Conan and E. Diamanti},
  journal={New Journal of Physics},
  volume={26},
  number={2},
  pages={023039},
  year={2024},
  publisher={IOP Publishing},
  note = {http://doi.org/10.1088/1367-2630/ad231c}
}

@article{theocharidis2025underwater,
  title={Underwater communication technologies: A review},
  author={T. Theocharidis and E. Kavallieratou},
  journal={Telecommunication Systems},
  volume={88},
  number={2},
  pages={54},
  year={2025},
  publisher={Springer},
  note = {http://doi.org/10.1007/s11235-025-01279-x}
}

@article{kaushal2016underwater,
  title={Underwater optical wireless communication},
  author={H. Kaushal and G. Kaddoum},
  journal={IEEE access},
  volume={4},
  pages={1518--1547},
  year={2016},
  publisher={IEEE},
  note       = {http://doi.org/10.1109/ACCESS.2016.2552538},
}

@book{lanzagorta2012underwater,
  title={Underwater communications},
  author={M. O. Lanzagorta},
  year={2012},
  publisher={Morgan \& Claypool Publishers},
  note       = {http://doi.org/10.1007/978-3-031-01678-3},
}

@article{ji2017towards,
  title={Towards quantum communications in free-space seawater},
  author={L. Ji and J. Gao and A.L. Yang and Z. Feng and X.F. Lin and Z.G. Li and X.M. Jin},
  journal={Optics Express},
  volume={25},
  number={17},
  pages={19795--19806},
  year={2017},
  publisher={Optical Society of America},
  note       = {http://doi.org/10.1364/OE.25.019795},
}

@article{zhao2019performance,
  title={Performance of underwater quantum key distribution with polarization encoding in ocean water},
  author={S.C. Zhao and X.H. Han and Y. Xiao and Y. Shen and Y.J. Gu and W.D. Li},
  journal={Journal of the Optical Society of America A},
  volume={36},
  number={5},
  pages={883--892},
  year={2019},
  publisher={Optical Society of America},
  note       = {http://doi.org/10.1364/JOSAA.36.000883},
}

@article{ata2025impact,
  title={Impact of Natural Turbulent Waters on Quantum Key Distribution: Temperature and Salinity Considerations},
  author={Y. Ata and K. Kiasaleh},
  journal={IEEE Journal of Oceanic Engineering},
  year={2025},
  volume={50},
  number={3},
  pages={2381-2393},
  publisher={IEEE},
  note = {http://doi.org/10.1109/JOE.2025.3551076}
}

@article{fung2006performance,
  title={Performance of two quantum-key-distribution protocols},
  author={C.H.F. Fung and K. Tamaki and H.K. Lo},
  journal={Physical Review A},
  volume={73},
  number={1},
  pages={012337},
  year={2006},
  publisher={APS},
  note = {http://doi.org/10.1103/PhysRevA.73.012337}
}

@article{elamassie2018performance,
  title={Performance characterization of underwater visible light communication},
  author={M. Elamassie and F. Miramirkhani and M. Uysal},
  journal={IEEE Transactions on Communications},
  volume={67},
  number={1},
  pages={543--552},
  year={2018},
  publisher={IEEE},
  note = {http://doi.org/10.1109/TCOMM.2018.2867498}
}

@article{ma2005practical,
  title={Practical decoy state for quantum key distribution},
  author={X. Ma and B. Qi and Y. Zhao and H.K. Lo},
  journal={Physical Review A},
  volume={72},
  number={1},
  pages={012326},
  year={2005},
  publisher={APS},
  note       = {http://doi.org/10.1103/PhysRevA.72.012326},
}

@article{ali2012practical,
  title={Practical {SARG04} quantum key distribution},
  author={S. Ali and S. Mohammed and M.S.H. Chowdhury and A.A. Hasan},
  journal={Optical and Quantum Electronics},
  volume={44},
  number={10},
  pages={471--482},
  year={2012},
  publisher={Springer},
  note = {http://doi.org/10.1007/s11082-012-9571-2}
  }

@article{fung2005performance,
  title={On the performance of two protocols: {SARG04} and {BB84}},
  author={C.H.F. Fung and K. Tamaki and H.K. Lo},
  journal={arXiv [quant-ph]},
  year={2005},
  note = {https://arxiv.org/abs/quant-ph/0510025}
}

@book{wilde2013quantum,
  title={Quantum information theory},
  author={M. Wilde},
  year={2013},
  publisher={Cambridge university press},
  note       = {http://doi.org/10.1017/CBO9781139525343},
}

@book{holevo2019quantum,
  title={Quantum systems, channels, information: a mathematical introduction},
  author={A. S. Holevo},
  year={2013},
  publisher={Walter de Gruyter GmbH \& Co KG},
  note       = {http://doi.org/10.1515/9783110273403},
}

\newpage

\begin{IEEEbiographynophoto}{Nour Rizk}
received the M.Sc. degree in Applied Mathematics and Modeling from the Université de Rouen Normandie - UFR Sciences et Techniques, Rouen, France, in 2023.
Since 2023, she has been pursuing the Ph.D. degree with the École Nationale Supérieure de Techniques Avancées (ENSTA, Institut Polytechnique de Paris), Brest, France, and the Lab-STICC, CNRS UMR 6285.
Her research focuses on quantum optical systems for underwater communication, quantum key distribution, and quantum information theory.
\end{IEEEbiographynophoto}

\begin{IEEEbiographynophoto}{Angélique Drémeau}
graduated from the IMT Atlantique (ex-ENSTB), Brest, France, in 2006. She received the Ph.D. degree from the University of Rennes, France, in 2010, and the HDR degree from the Université de Bretagne Occidentale (UBO), Brest, France, in December 2022.
She is currently a Full Professor at the École Nationale Supérieure de Techniques Avancées
(ENSTA, Institut Polytechnique de Paris), Brest. Her research interests include the information theory and wave propagation in complex environments. 
\end{IEEEbiographynophoto}

\begin{IEEEbiographynophoto}{Arnaud Coatanhay}
graduated from the École Supérieure d’Ingénieurs en Génie Electrique (ESIGELEC), Rouen, France, in 1993. He received the Ph.D. degree from the University of Le Havre,
Le Havre, France, in 2000, and the HDR degree from the Université de Bretagne Occidentale (UBO), Brest, France, in June 2022.
He is currently a Full Professor at the École Nationale Supérieure de Techniques Avancées
(ENSTA, Institut Polytechnique de Paris), Brest. His research interests include the modeling of electromagnetic wave propagation in complex and random media. For almost ten years, his research has been devoted more specifically to quantum optics in natural environments, with applications in telecommunications and remote sensing.
\end{IEEEbiographynophoto}

\vspace{11pt}

\vfill
\end{document}